\documentclass[aps,prl,reprint,superscriptaddress,longbibliography]{revtex4-1}
\usepackage{blindtext}
\usepackage{graphicx}
\usepackage{amsmath,amssymb}
\usepackage{multirow}
\usepackage{url}
\usepackage{color,soul}
\usepackage[colorlinks,
            linkcolor=blue,
            anchorcolor=blue,
            citecolor=blue,
            ]{hyperref}
\usepackage{setspace}

\begin{document}

\title{Time Domain Generalization of the Random Coupling Model and Experimental Verification in a Complex Scattering System}

\author{Shukai Ma}
\email{skma@umd.edu}
\affiliation{Quantum Materials Center, Department of Physics, University of Maryland, College Park, Maryland 20742, USA}
\author{Thomas M. Antonsen}
\affiliation{Department of Physics, University of Maryland, College Park, Maryland 20742, USA}
\affiliation{Department of Electrical and Computer Engineering, University of Maryland, College Park, Maryland 20742-3285, USA}
\author{Steven M. Anlage}
\affiliation{Quantum Materials Center, Department of Physics, University of Maryland, College Park, Maryland 20742, USA}
\affiliation{Department of Electrical and Computer Engineering, University of Maryland, College Park, Maryland 20742-3285, USA}

\begin{abstract}

Electromagnetic (EM) wave scattering in electrically large, irregularly shaped, environments is a common phenomenon. 
The deterministic, or first principles, study of this process is usually computationally expensive and the results exhibit extreme sensitivity to scattering details. 
For this reason, the deterministic approach is often dropped in favor of a statistical one.  
The Random Coupling Model (RCM) \cite{Alt1998} is one such approach that has found great success in providing a statistical characterization for wave chaotic systems in the frequency domain. 
Here we aim to transform the RCM into the time domain and generalize it to new situations. 
The proposed time-domain RCM (TD-RCM) method can treat a wave chaotic system with multiple ports and modes. 
Two features are now possible with the time-domain approach {for chaotic resonators}: the incorporation of early-time short-orbit transmission path effects between the ports, and the inclusion of arbitrary nonlinear or time-varying port load impedances. 
We have conducted short-pulse time-domain experiments in wave chaotic enclosures, and used the TD-RCM to simulate the corresponding experimental setup. 
We have also examined a diode-loaded port and compared experimental results with a numerical TD-RCM treatment and found agreement.

\end{abstract}

\maketitle

\begin{spacing}{1}

\section{I. Introduction}

The complicated scattering of electromagnetic (EM) waves can be found in many applications \cite{Alt1998, Dietz2006, Kaya2009, Tait2011, Ma2019}.
Examples include the propagation of WiFi signals in an office space, and the propagation of high-power microwave signals in interconnected cabins in ships and aircraft. 
When the enclosure characteristic dimension is much larger than the operating wavelength, the propagation of waves can be thought of in terms of ray trajectories. 
In this limit, the ray dynamics are extremely sensitive to geometrical details, as well as to initial conditions. 
If the trajectories exhibit chaos, that is, nearby trajectories diverge from one and other exponentially upon undergoing many bounces, and this occurs before a  wave packet following the trajectory damps out, such enclosures are labeled as wave-chaotic. 
 
The complicated dynamics of EM waves pose significant challenges for the exact solution of the EM fields. 
With powerful computers, accurate modeling of wave propagation can be simulated, but only when the exact geometrical details of the scattering system are known \cite{Taflove1988, Yang1997, Davies2009}. 
However, the precise geometrical details of such large systems are usually unknown in practice. 
Moreover, the results computed from deterministic methods will be rendered invalid when perturbations are applied to the system, for example, by a simple change of system boundary. 
Under these circumstances, statistical approaches are more suitable for studying wave-chaotic systems.

Random Matrix Theory (RMT) is widely applied to describe the statistics of spectral properties of wave chaotic systems. 
Researchers originally developed RMT to model the spectra of the complicated Hamiltonians of large nuclei \cite{Wigner1955,Dyson1962,Mehta2004}.
Later,  Bohigas, Giannoni, and Schmit (BGS) proposed the applicability of RMT to simple wave propagation in domains with complicated boundaries.  
It is now surmised that all systems that show ray-chaos in the classical regime will have wave properties that share universal statistical properties \cite{Bohigas1984}.
Based on the BGS conjecture, RMT has found broad application in various physical systems (atoms, quantum dots, quantum wires, microwave enclosures, acoustic resonators, and others) \cite{Haq1982,stockmann2000quantum,Alhassid2001,Tanner2007,haake2010,Auregan2016}.
Depending on certain basic symmetries, the spectral properties of a specific chaotic system can be categorized into one of three groups \cite{Mehta2004,haake2010}. 
The first two groups are the Gaussian Orthogonal Ensemble (GOE) and Gaussian Unitary Ensemble (GUE) of random matrices for a system with and without time-reversal symmetry, respectively. 
The third group is the so-called Gaussian Symplectic Ensemble (GSE) of random matrices for a system with half-integer spins and time-reversal symmetry.

{Treatments of the scattering property statistics of wave chaotic systems have been carried out for some time} \cite{RN22957, stockmann2000quantum, RN10222, RN363, Fyodorov2005, Mitchell2010, Fyodorov2010, RN23593, RN27431, Nock2014, Dietz2015, Fyodorov2015, Schomerus2017, Hemmady2005,Zheng2006,Zheng2006a}.
The Random Coupling Model (RCM), developed for electromagnetic applications \cite{Hemmady2005,Zheng2006,Zheng2006a}, has also found great success in characterizing the statistical properties of various frequency-domain quantities, for example, the full scattering (S -) matrix and the impedance (Z -) matrix of wave chaotic systems \cite{Hemmady2012,Gradoni2015}.

\subsection{I. Legacy Random Coupling Model and Short-Orbit Treatment}

To use the RCM one must first identify the ports of entry and exit for waves in the enclosure.  
The RCM then provides a statistical expression for the impedance matrix that linearly relates the port currents and voltages.  
For a chaotic cavity, a model for the fluctuating impedance matrix was given by Hart \cite{Hart2009}, based on the work of Brouwer \cite{Brouwer1995}. The model expresses the statistically varying impedance matrix $\underline{\underline{Z}}_{cav}$ in terms of an average matrix and a fluctuating matrix,
\begin{equation} \label{eq:tma1}
\underline{\underline{Z}}_{cav}=i\underline{\underline{X}}_{avg} + \underline{\underline{R}}_{avg}^{1/2}\cdot \underline{\underline{\xi}}\cdot  \underline{\underline{R}}_{avg}^{1/2}. 
\end{equation}
Here $\underline{\underline{Z}}_{avg}(\omega)=\underline{\underline{R}}_{avg} + i\underline{\underline{X}}_{avg}$ is a frequency or realization averaged impedance matrix and $\underline{\underline{\xi}}$ is a statistically fluctuating matrix drawn from the Lorenzian ensemble introduced by Brouwer \cite{Brouwer1995}.
The average impedance matrix describes the early-time dynamics of fields in the enclosure.  This includes radiation from the port into the enclosure as well as the propagation of radiation from one port to another in the enclosure over short paths. 
Hart \cite{Hart2009} showed that in the short wavelength limit the average impedance matrix could be expressed as the sum of the diagonal radiation impedance matrix, which characterizes the coupling of the ports to the enclosure, and a second matrix containing both diagonal and off-diagonal contributions between the ports from the short paths,
\begin{equation} \label{eq:tma2}
\underline{\underline{Z}}_{avg}=\underline{\underline{R}}_{rad} + i\underline{\underline{X}}_{rad} + \underline{\underline{R}}_{rad}^{1/2}\cdot \underline{\underline{\zeta}}\cdot  \underline{\underline{R}}_{rad}^{1/2}. 
\end{equation}
Here $\underline{\underline{Z}}_{rad}=\underline{\underline{R}}_{rad} + i\underline{\underline{X}}_{rad}$ is the diagonal radiation impedance and the matrix $\underline{\underline{\zeta}}$ can be expressed in the short wavelength limit as a sum over paths going from one port to another \cite{Hart2009}, 
\begin{equation} \label{eq:tma3}
\underline{\underline{\zeta}} = \sum_{paths} \underline{\underline{C}}_{path} exp[i(k + i\kappa) L_{path} - i\pi/4].
\end{equation}
In Eq. \ref{eq:tma3}, $\underline{\underline{C}}_{path}$ gives the strength of the path’s contribution, $k = \omega/c$ where $c$ is the propagation speed, $\kappa$ is the attenuation rate, and $L_{path}$ is the length of the path.

The length of the retained paths depends on how the average is taken \cite{Hart2009, Yeh, Zhou2017, Xiao2018, Chen2022}. 
If the average is over a window of frequencies, then only paths with travel times shorter than the inverse of the frequency window survive the averaging process.  
If the average is taken by varying the orientation or position of a mode stirrer, then only paths that do not intersect the stirrer survive the averaging process \cite{Hart2009}.  
Evaluation of Eq. \ref{eq:tma3} involves identifying the relevant short paths and computing the coefficients and path lengths \cite{Yeh}.  
Alternatively, the average impedance matrix can be evaluated by measuring or computing the early time impulse response of the enclosure \cite{Zheng2006a, Zheng2006b}, multiplying it by a decaying exponential function of time, and Fourier transforming the product.  
This gives a frequency window average of the impedance matrix, where the width of the window is the exponential decay rate.

Equation. \ref{eq:tma1} describes the realization-to-realization fluctuations in a system’s impedance matrix but does not model the frequency dependence of a single realization’s impedance. 
To do this we first need to characterize the processes that contribute to the impedance matrix and the time scales over which they operate.  The diagonal radiation impedance characterizes the entry of wave energy into the enclosure through the ports.  This will vary in frequency on a scale based on the size of the port $L_{port}$, $\Delta \omega_{port} = c/L_{port}$.  
It is expected this is the largest frequency scale in the problem.  
The average impedance varies on the port frequency scale, but also has finer frequency variations on the scale $\Delta \omega_{path} = c/L_{path} \ll \Delta \omega_{port}$ if short orbits are retained.  
Finally, we come to the finest frequency scale in the problem, which will be attributed to the fluctuating matrix  $\underline{\underline{\xi}}$.  
This finest scale will be set by the properties of the resonant modes of the enclosure.  It will generally be either the average spacing between modes, $\Delta \omega$, or the mode Q-width, $\omega / Q$, whichever is larger.  The ratio of these two frequency scales is labeled the universal loss parameter (defined below) and it characterizes the distribution of values of the elements of the matrix $\underline{\underline{\xi}}$.  
The average spacing and quality factor vary over the frequency range of the port $\Delta \omega_{port}$.  
We will soon specialize in the case of a band-limited signal centered around a carrier and define these quantities at the carrier frequency.

Given the range of frequency and time scales it is generally difficult to accurately simulate a system on all three scales at the same time.  
Thus, we specialize to a situation where the excitation signal has a relatively narrow bandwidth $\Delta \omega_{BW}$ around a central or carrier frequency $\omega_c$, $\Delta \omega_{BW} < \omega_c$.  
We will assume that the bandwidth is comparable to the path frequency $\Delta \omega_{BW} \approx \Delta \omega_{path} \ll \omega_{port}$.  
Thus, we will construct a model which describes the response of the enclosure to a pulse envelope that has time variations on the scale of the length of the short orbits and longer, and which modulates a carrier.  In this approximation the time scales associated with the injection of power through the ports will be treated as instantaneous, and the radiation impedance matrix will be evaluated at the carrier frequency, $\omega_c$.  
We will make an additional approximation that the cavity loss factor is high ($\omega / Q \gg \Delta \omega$, where $\Delta \omega$ and $Q$ are now specified to be the average mode spacing and quality factor for frequencies near the carrier) and that the short orbit contributions to the impedance matrix are small $|\underline{\underline{\zeta}}| \ll 1$.  
What these approximations mean in practice is that only a small fraction of the energy injected into the enclosure escapes through a labeled port.  
In this limit by combining Eqs. \ref{eq:tma1} and \ref{eq:tma2} the cavity impedance matrix takes a simple form in which the short orbit effects and wave chaotic mode effects are additive,
\begin{widetext}
    \begin{equation} \label{eq:tma4}
    \underline{\underline{Z}}_{cav}=i\underline{\underline{X}}_{rad}(\omega_c) + \underline{\underline{R}}_{rad}^{1/2}(\omega_c) \cdot [\underline{\underline{\zeta}}(\omega) + \sum_{n-range} \underline{\underline{\xi}}_n(\omega)]\cdot  \underline{\underline{R}}_{rad}^{1/2}(\omega_c). 
    \end{equation}
\end{widetext}

Here the random matrices $\underline{\underline{\xi}}_n$ are continuous, frequency-dependent, matrices that fluctuate from realization to realization and provide the same statistical properties at a single frequency as the matrix given in Eq. \ref{eq:tma1},
\begin{equation} \label{eq:tma5}
\underline{\underline{\xi}}_n(\omega) = - \frac{2i\omega}{\pi} \frac{\Delta \omega \underline{c} \, \underline{c}^T}{\omega_n^2 - i \omega\omega_c/Q - \omega^2}.
\end{equation}
In Eq. \ref{eq:tma4} the sum is over a group of eigenfrequencies of the closed cavity, $\omega_n$, which we replace by a spectrum based on RMT (Appendix A). 
Further, we place this group of eigenfrequencies over a range whose center is the carrier frequency, and whose extent is larger than the Q-width of the cavity and the bandwidth of the signal, allowing the sum to converge.  
In this way the average contribution of the sum to the imaginary part of Eq. \ref{eq:tma4} is negligible as the average part is already included in $\underline{\underline{X}}_{rad}(\omega_c)$.  
In Eq. \ref{eq:tma5}, we have now defined the mean spacing in frequency between modes with frequency near $\omega_c$, $\Delta \omega = \langle \omega_{n+1} - \omega_n \rangle$, the average quality factor for modes with frequency near $\omega_c$, Q, and a column vector $\underline{c}_n$ of dimension $M$ (the number of ports) whose elements are independent and identically distributed zero mean, unit variance, Gaussian random variables.  We also introduce the previously mentioned loss parameter which is the ratio of the Q-width of the mode resonances to the average mode spacing,
\begin{equation} \label{eq:tma6}
\alpha = \omega_c / (Q \Delta \omega).
\end{equation}
A detailed derivation of this result for the case in which $\underline{\underline{Z}}_{avg} = \underline{\underline{Z}}_{rad}$, the diagonal radiation impedance, is presented in Appendix B.

We can evaluate the expected value of $\underline{\underline{Z}}_{cav}$ in the weak damping, dense spectrum limit $Q \gg \alpha \gg 1$, by noting $\langle \underline{c} \, \underline{c}^T \rangle = \underline{\underline{I}}$ , the identity matrix, and by replacing the sum over modes with an integral over mode frequency.  
We find $\sum_{n-range} \Delta \omega \underline{\underline{\xi}}_n \rightarrow \int \underline{\underline{\xi}}_n d\omega_n = \underline{\underline{1}}$, and consequently agreement with Eq. \ref{eq:tma2}.

The frequency dependence of the impedance matrix has consequences for the time dependence of signals in our model system.  First, we note that all the frequency dependence is contained in the square bracket term in Eq. \ref{eq:tma4}. 
This term governs the propagation of signals between the ports.  It consists of two terms, a short orbit contribution and a fluctuating chaotic mode contribution. We note that the average impedance matrix, Eq.\ref{eq:tma2}, exhibits prompt reflections from the excited ports and appropriately delayed reflections and transmission to other ports due to inclusion of short orbits.

Although the average response will respect all time delays, individual realizations described by Eqs.\ref{eq:tma4} and \ref{eq:tma5} will not.  As we will describe in coming sections, Eq. \ref{eq:tma5} will be realized by a system of second order differential equations in time for the individual "chaotic" mode amplitudes.  The response of the mode amplitudes to steady excitation is to initially grow linearly in time and then come into equilibrium on the time scale of the quality factor $Q/\omega_c$.  
This initial growth will cause some low level, fluctuating transmitted signals to appear prematurely.  
The error being of order $\omega_c T_{path} /Q = T_{path} / T_Q$, where $T_{path}$ is the travel time for signals on a path connecting the ports and $T_Q$ is the decay time for the fields in the enclosure.  For a reverberant enclosure the time of flight for signals between ports is much less than the decay time, $T_{path} / T_Q \ll 1$.

If the injected signal is for communications purposes the symbol rate, $T_s^{-1}$, becomes important.  One can expect that paths with time durations $T_{path} / T_s \leq 1$ can contribute coherently to the transmission, while the contributions from longer paths will cause inter-symbol intereference. 
In this case one would want to include in the average impedance the contributions from paths with sufficiently short durations $T_{path} / T_s \leq 1$, while the statistically fluctuating response models the interference from previously injected signals with delayed arrival times.


{Thus, in our model the signal arriving at a port is the sum of two terms as represented in the square bracket in} Eq. \ref{eq:tma4}.  
{The first term is an average, ``or coherent'', portion containing the short orbit contributions, and which satisfies all physical time delays. 
The second term is the stochastic contribution expressed in terms of mode amplitudes. 
The amplitude of a mode will be determined by solving a second-order differential equation in time with the input currents at the port as sources.} 
{If one thinks about expressing the solution of this differential equation as a convolution, the kernel has a time duration $Q/\omega_c=T_Q \gg T_{path}$. 
Thus, in the highly reverberant cavities that we consider, the major contributions to the stochastic mode amplitudes will come from signals injected well before the current time, and that have reverberated in the cavity for a time $T_Q$. 
Thus, causality will effectively be satisfied.}


In this paper, we aim to recast the RCM from the frequency domain to the time domain. 
Previous work has directly applied the frequency-domain RCM to compute the system time-domain quantities (reflected voltage at the input port) by Fourier transform \cite{Hart2009}.
However, this method can not be used to describe nonlinear devices in the scattering system. 
{The proposed time domain RCM (TD-RCM) method is a true time-domain calculation, in the sense that quantities evolve in the time domain either as mode amplitudes or as convolutions. }
Because the quantities are evolved in time, one can treat various types of port loads including both linear and nonlinear, passive and active devices, and time-dependent loads, such as those associated with time-dependent logic states of electronics \cite{Ma2021piers,Chen2022}.
The TD-RCM also explicitly includes non-universal effects through the incorporation of the port radiation impedances, losses, and short orbit effects.

The time domain treatment presented here can be compared with previous approaches \cite{tomsovic1993long, Castaldi2012}.  One approach, known as Statistical Energy Analysis (SEA), is to calculate the time dependence of the spatially-averaged energy density in one or more coupled reverberation chambers \cite{Tait2011}. The resulting system consists of a small number of coupled ordinary differential equations, one per cavity or enclosure, with coupling coefficients describing the injection of energy in each enclosure and its leakage from one enclosure to another.  A more refined approach, known as Dynamical Energy Analysis (DEA), is to solve a kinetic equation for the transport of wave action through phase space in complex  structures \cite{weaver2006transport, tanner2009dynamical}. In both these approaches the quantities solved for are quadratic in the fields and averaged over the time period of the oscillating fields.  Thus, the effects of interference are neglected, while they are  retained in the treatment presented here.  A third statistical time-domain method is based on the Baum-Liu-Tesche equations \cite{Tesche2007} which model communication among enclosures in terms of transmission lines. 
Other statistical time-domain methods are based on multipath ray sums \cite{saleh1987statistical,rappapor1989} where the arrival times are Poissonian, the amplitudes are governed by Rayleigh statistics, and the phases are assumed random.

{There is interesting related work in the rapidly developing area of reconfigurable intelligent surfaces RIS }\cite{Frazier2020, Wang2022}. 
{Here researchers are interested in modeling and controlling the distribution and redistribution of field amplitudes in reverberant enclosures containing programable meta-surfaces.  These surfaces, consisting of an array of reflecting cells, scatter incident waves in different directions depending on the state of the array, which is electrically controlled.  As a result of the controlled scattering, it is possible to redirect waves and create hot and cold spots at arbitrary points in the enclosure.  
An application of such an RIS is to enhance wireless communications in a reverberant environment }\cite{DelHougne2018}.
{Given this application, a time dependent description of the scattering and interference of the signals following different paths is desirable.  As we have discussed, the model  we propose includes two types of path by which signals are transported between ports: deterministic paths as contained in the average impedance matrix, and statistical paths as contained in the random matrices Eq.(5). Application of our model to the RIS communication problem would require including the RIS in the short orbit portion of the average impedance matrix where precise time of flight information is retained, and treating the reverberant background of signals interfering with the communications signal statistically.}


The outline of the paper is as follows. 
We first introduce the fundamental equations of the TD-RCM method in section II. 
Section III introduces the short-pulse experimental verification of TD-RCM theory, followed by a comparison between the experimental results and the TD-RCM simulated results in sections IV and V. 
We give an example of the use of TD-RCM for nonlinear ports in the context of reservoir computing in Section VI.
We summarize the results in section VII.

\section{II. TD-RCM formulation}

In this section, we discuss the formulation of the TD-RCM method. 
We first present the basic TD-RCM statistical model in Section II.A. 
We next describe the treatment of system port loads (both linear and nonlinear loads) in Section II.B. 
In Section II.C, we present a method to include short-orbits (SO) in TD-RCM, thus incorporating the non-statistical aspects of the early-time transmission inside the cavity.

\begin{figure}
\centering
\includegraphics[width=0.5\textwidth]{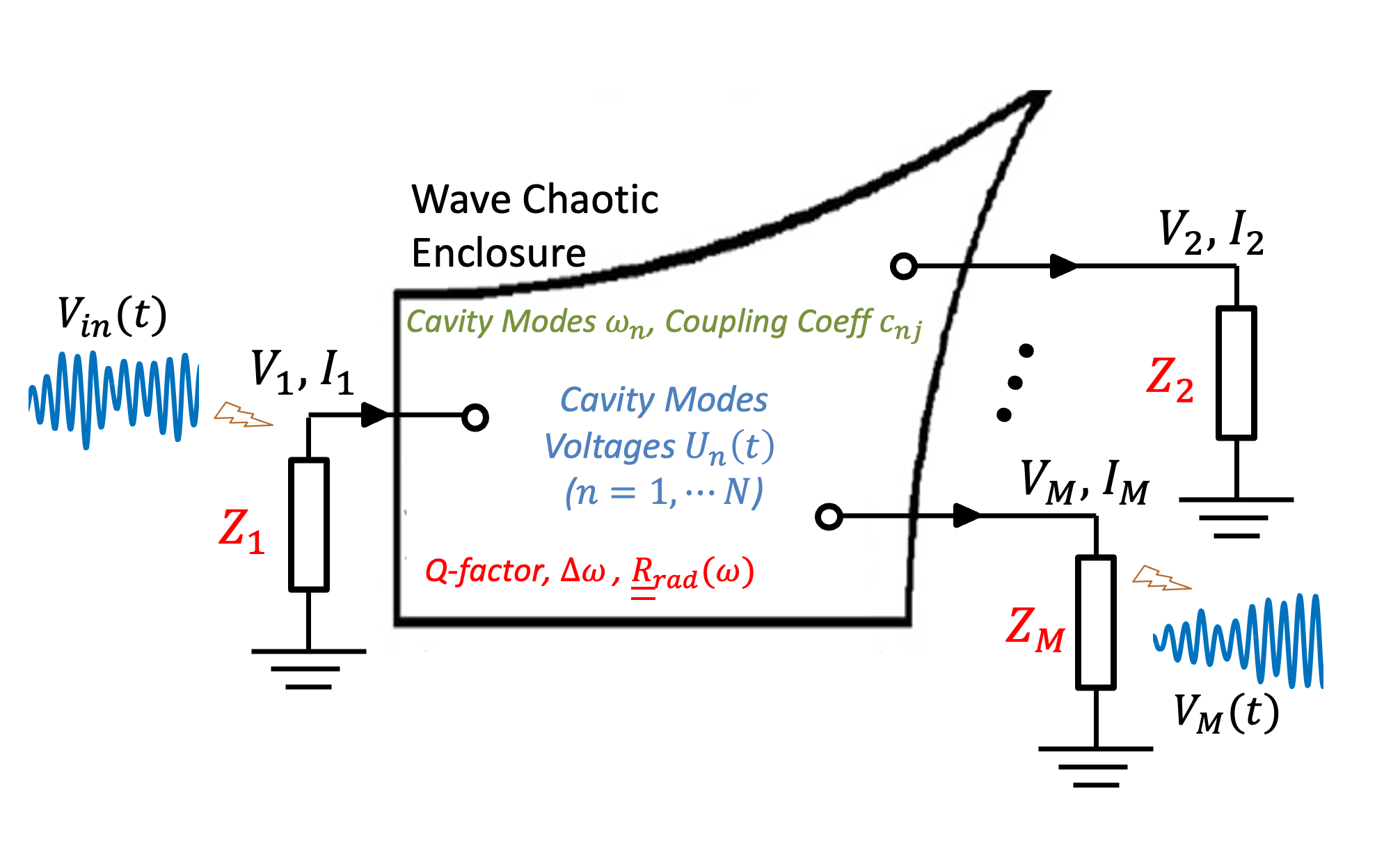}
\caption{\label{fig:outline} 
Schematic summary of the TD-RCM method.
The injected waveform $V_{in}(t)$ is sent into a wave chaotic enclosure through port 1.
The enclosure has $N$ eigenmodes excited.
The TD-RCM model computes the voltage and current on all $M$ ports, as well as all $N$ cavity mode voltages.
The system-specific quantities (cavity Q-factor, mean-mode-spacing $\Delta \omega$, the radiation resistance $\underline{\underline{R}}_{rad}(\omega)$ of the ports, and the load impedances attached to the ports, $Z_j$) are color-coded in red, while the RMT generated quantities (cavity mode frequencies) are colored in green.
The result is predictions for the voltage waveforms appearing on all ports as a function of time, $V_i(t), i = 1, \dots, M$.
For example, a sketch of the voltage wave prediction at port $M$ is shown as $V_M(t)$ in the figure.
}
\end{figure}

\subsection{II.A TD-RCM Evolution Equations}

The time-domain Random Coupling Model (TD-RCM) models an enclosure supporting $N$ eigenmodes covering a range of frequencies such that the integral in Eq.(5) converges.
The enclosure can have $M$ ports, where each of the ports can serve as a transmitting (TX) port or a receiving (RX) port, or both.
We use Fig. \ref{fig:outline} to summarize the variables and parameters of the TD-RCM method.
The TD-RCM computes the temporal evolution of cavity modes and port variables. 
The type of load connected at each of the ports can be different, and the degree of lossyness for each of the modes can also be specified.
The radiation resistance of the port $R_{rad}$ characterizes the free-space radiation property of the port antenna \cite{Hemmady2005, Zheng2006, Zheng2006a}.
The cavity modes ($U_n, n = 1, 2, \cdots, N$) are represented by the real voltages $U_n$, where $n$ is the mode index, and represents the amplitude of excitation of the $n^{th}$ eigenmode of the closed system.
The signal at port $j$ is represented by the port voltage $V_j$ and port current $I_j (j = 1, \dots, M)$.

Each cavity mode is described by a driven damped harmonic oscillator equation,
\begin{equation} \label{eq:modeV}
    \frac{d^2}{dt^2} U_n + v_n \frac{d}{dt} U_n + \omega_n^2 U_n = \omega_n \sum_{j=1}^M c_{nj} K_{nj} \frac{d}{dt} I_j, 
\end{equation}
where port subscript $j$ runs through all $M$ system ports.
The port voltages ($V_j, j = 1, 2, \cdots, M$) are modeled by a linear summation over all cavity modes,
\begin{equation} \label{eq:portV}
    V_j = \sum_{n=1}^N c_{jn} \sqrt{ \frac{\epsilon}{\mu}} K_{nj} U_n - V_{jc}.
\end{equation}
The quantity $V_{jc}$ represents the electrostatic mode contribution of port $j$, defined as 
\begin{equation}
C_p \frac{d V_{jc}}{dt} = -I_j.
\end{equation}
The term $C_p$ is a port capacitance that is important at low frequencies for dipole-like ports.
In Eq. \ref{eq:modeV}, the damping term for a system mode is defined as $v_n =\omega_n/Q_n$, where we allow each mode to have a different quality factor.
The quantities $\omega_n$ and $\Delta \omega_n$ are the eigenmode frequency and mean-mode-spacing near $\omega_n$.

As in the frequency-domain RCM, the quantity $c_{jn}$ is a normalized, fluctuating coupling coefficient between mode $n$ and port $j$ (assumed to be Gaussian random variables).
The factor 
\begin{equation}  \label{eq:knj}
K_{nj} = - \left( \frac{\mu}{\epsilon} \right )^{\frac{1}{4}} \left[ \frac{2R_{rad,j}(\omega_n) \, \Delta\omega_n}{\pi\omega_n } \right]^{\frac{1}{2}}
\end{equation}
is a dimensional factor having units of Ohms that give the strength of the coupling of mode $n$ to port $j$ (see the derivation in Appendix B).
The quantities $\mu$ and $\epsilon$ are the (assumed uniform) permeability and permittivity of the enclosure volume material. 
The quantity $R_{rad,j}(\omega_n)$ is the frequency domain radiation resistance of port $j$ evaluated at frequency $\omega_n$.
In Eq. \ref{eq:portV}, the random coupling coefficient $c_{jn}$ and mode voltages $U_n$ are the same quantities as in Eq. \ref{eq:modeV}.
In the analysis presented here we modify Eq.\ref{eq:tma4} slightly in that we move the factors $R^{1/2}_{rad}(w_c)$ inside the sum on modes-n, and replace the carrier frequency, $\omega_c$, the argument of the radiation resistance with the mode frequency, $\omega_n$.  
In the narrow bandwidth limit, this is a minor change.   
However, it more accurately represents the mode expansion analysis in Appendix B.

For an enclosure with $N$ modes and $M$ ports, we have $N$ versions of Eqs. \ref{eq:modeV} and $M$ versions of Eqs. \ref{eq:portV}.
The unknown quantities that need to be solved are the $N$ $U_n$s, the $M$ currents $I_j$, and the $M$ voltages $V_j$, all of which are functions of time.
Usually we set the initial condition ($t = 0$) as $U_n = dU_n/dt = V_{jc} = 0$ for all $n$ and $j$, representing a quiet cavity.
When the input pulses are turned on, the port voltage values at the transmitting (TX) ports will have nonzero values as well as nonzero time derivatives. 
As shown in the RHS of Eq. \ref{eq:modeV}, the change of port signals will then drive cavity modes to oscillate.
The oscillating cavity modes will further contribute to the nonzero signals at all ports (the RHS of Eq. \ref{eq:portV}).
In addition to Eqs. \ref{eq:modeV} and \ref{eq:portV}, one needs the explicit relationship between the port voltage and current signals that describes the load at the port to solve for all system mode ($U_n$) and port quantities ($I_j, V_j$).

\begin{figure}
\centering
\includegraphics[width=0.5\textwidth]{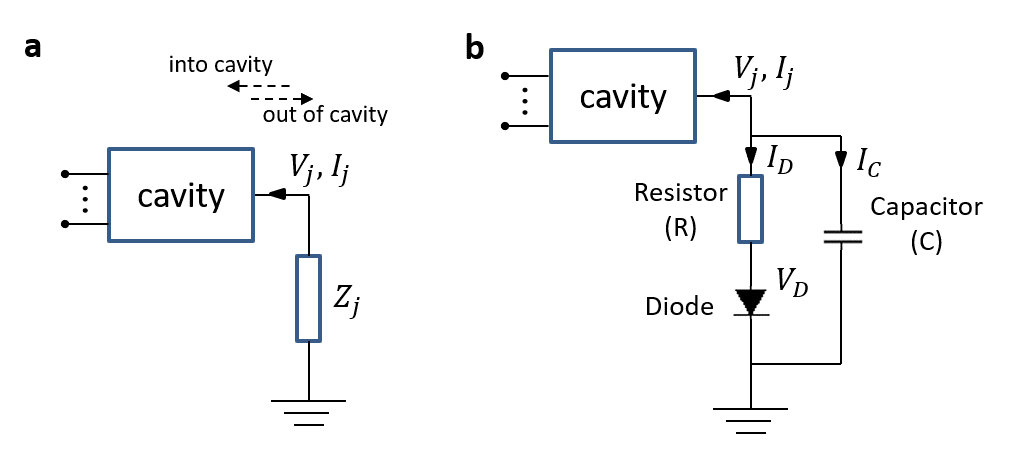}
\caption{\label{fig:portload} 
Diagrams of the TD-RCM port load modeling.
\textbf{a.} The linear load at port $j$ is represented with an impedance $Z_j$.
The dashed arrows mark the direction of waves flowing into and out of the cavity.
\textbf{b.} Schematic of a diode-loaded port with the resistor, diode, and capacitor components labeled. 
}
\end{figure}

\subsection{II.B System Port Treatments}

In a time domain model the equations for the mode amplitudes $U_n$, Eq. \ref{eq:modeV}, must be integrated in time along with the dynamic variables describing the ports. 
The source term in Eq. \ref{eq:modeV} is proportional to the time derivative of the port current, while the port voltage is linearly related to the mode amplitudes through Eq. \ref{eq:modeV}.  
These variables, port voltages, and port currents are dynamically related by the models describing the port loads.  
In this subsection, we discuss the treatment of two generic types of loads that are present at ports: a linear impedance load (Fig. \ref{fig:portload} (a)) and a nonlinear load (Fig. \ref{fig:portload} (b)), where we utilize the specific example of a diode-loaded port.
We first show the treatment of ports with a linear load.

If the port is excited by an incoming wave on a transmission line, with characteristic impedance $Z_j$, the voltage and current at the port $j$ can be related by
\begin{equation} \label{eq:linearload}
\begin{aligned}
    V_j &= V_{in} + V_{out}, \\
    I_j &= (V_{in} - V_{out})/Z_j,\\
\end{aligned}
\end{equation}
where $V_{in}$ ($V_{out}$) represents the incident (reflected) voltage wave coming into (out of) the cavity.
The direction of the waves is shown by the dashed arrows in Fig. \ref{fig:portload}, so that ``in'' corresponds to waves going into the cavity, and ``out'' represents the waves that leave the cavity.
If the port is simply connected to a load, $Z_j$, then we set $V_{in} = 0$.
The value of the load impedance $Z_j$ (often 50 $\Omega$) is represented by the load connected between the cavity and the ground in Fig. \ref{fig:portload} (a).
Note that the TD-RCM allows the cavity to have multiple ports, shown schematically in Fig. \ref{fig:portload} with the lines connected to the left side of the cavity.
For a signal input port, we have $I_j = (V_{in} - V_{out})/Z_j = (2V_{in} - V_{j})/Z_j$ where $V_{in}$ is the waveform injected into the cavity at that port.
For an output port, we have $I_j = -V_j/Z_j$ since $V_{in} = 0$.

One may express the I-V relation of both input and output ports by the following vector expression
\begin{equation} \label{eq:linearload2}
    \underline{I} = \underline{\underline{Y}} \, \underline{V} + \underline{D}.
\end{equation}
Vector quantities are used here for simplicity: $\underline{I}$ and $\underline{V}$ are two M-by-1 column vectors for all port currents and voltages, whose elements are $I_j$ and $V_j$ $(j = 1, \dots, M)$.
Here $\underline{\underline{Y}}$ is a diagonal M-by-M matrix where $Y_{jj} = -1/Z_{j}$ and $\underline{D}$ is a time-dependent vector with $D_j = 2V_{in, j}/Z_j$.
With Eqs. \ref{eq:modeV}, \ref{eq:portV} and \ref{eq:linearload2}, we have now obtained enough equations to solve for the port voltages.
Note that the values of port load impedance $(Z_j)$ and the injected voltage signals $(V_{in, j})$ values are treated as known quantities.
In a finite difference implementation, assuming $U_n$ and $dU_n/dt$ are known at time $t$, then $V_j$ and $dV_j/dt$ are known through Eq. \ref{eq:portV}.
Using Eq. \ref{eq:linearload2} then gives $dI_j/dt$ and consequently, Eq. \ref{eq:modeV} can be used to update $dU_n/dt$ and $U_n$.

We next study the case with nonlinear output ports and use a generic diode-loaded port as a representative example, as illustrated in Fig. \ref{fig:portload} (b). 
(Note that the TD-RCM can also treat nonlinear input ports utilizing this same formalism.)
{As a finite-difference-method-based time-domain model, one of the strengths of TD-RCM is the ability to treat ports loaded with nonlinear devices.
Simulating nonlinear components is a unique advantage of true time-domain methods as opposed to the methods that compute time-domain signals from frequency-domain quantities} \cite{Hart2009, Faqiri2022}.
As shown in Fig. \ref{fig:portload} (b), the diode-loaded port is modeled with a series resistance ($R$) and a shunt capacitance ($C$). The diode-port is modeled by the following equations:
\begin{equation} \label{eq:diode}
\begin{aligned}
    V_j &= V_D + R \, I_D, \quad I_j = -(I_C + I_D) \\
    I_C &= C \frac{d}{dt} V_j, \quad I_D = I_0[exp(\frac{qV_D}{k_BT})-1].
\end{aligned}
\end{equation}
The quantities $V_D, I_D$, and $I_C$ are the voltage drop at the diode, the currents at the diode path, and the shunt capacitor path (Fig. \ref{fig:portload} (b)).
The diode parameters $I_0, q, k_B, T$ are the reverse current, elementary charge, Boltzmann constant, and the temperature in Kelvin.
The subscripts $D$ and $C$ represent the diode and capacitor. For convenience, we set $I_D = 0$ for negative voltages. Eq. \ref{eq:diode} can be simplified to 
\begin{equation} \label{eq:diode2}
    C \frac{dV_j}{dt} + \frac{V_j}{R} + I_j = -\frac{V_D}{R},
\end{equation}
\begin{equation} \label{eq:diode3}
    V_D = \frac{k_BT}{q} ln(1 + \frac{V_j - V_D}{R \, I_0}).
\end{equation}
The appearance of $dV_j/dt$ in Eq. \ref{eq:diode2} now adds a complication to the finite difference solution of our system of equations.
In particular, if Eq. \ref{eq:diode2} is differentiated in time to express $dI_j/dt$, and $dI_j/dt$ is inserted in Eq. \ref{eq:modeV}, then the second order derivative of $V_j$ appears on the right side of Eq. \ref{eq:modeV} and the second derivative of $U_n$ appears on the left side.
These must be solved for consistently with the relation Eq. \ref{eq:portV}.
This can be done by introducing a predictor-corrector method.
The voltage across the diode $V_D$ is computed by solving the transcendental equation \ref{eq:diode3}. 
We examine the effectiveness of TD-RCM modeling of nonlinear elements in Appendix C, and provide an example of the TD-RCM modeling of a reservoir computer based on a microwave cavity containing diode-loaded output ports in Section VI.

\subsection{II.C Inclusion of System-Specific Short-Orbits}

So far we have discussed the treatment of the statistically fluctuating contribution to the impedance matrix Eq. \ref{eq:tma4}. 
We now turn to the average contribution contained in the matrix $\underline{\underline{\zeta}}$, Eq. \ref{eq:tma3}.
Including this term gives us the ability to describe the propagation of waves directly from port to port, or indirectly with only a few reflections from the enclosure’s walls.  
This is referred to as the short-orbit contribution.  
Here, rather than computing $\underline{\underline{\zeta}}$ as a sum over paths as given by Eq. \ref{eq:tma3} as was done in Refs. \cite{Hart2009, Yeh}, we will assume that it can be measured.

{We define short orbits as system-specific trajectories that link ports by means of either ballistic or few bounce, ray orbits.
Short orbits can persist in ensemble measurements in which one or more perturbers are moved to block various orbits that connect two ports of a complex scattering system }\cite{Hart2009}.  
{Such short orbits have been accounted for in frequency-domain treatments of complex scattering by a number of authors }\cite{Baranger1996, Bulgakov2006, Hart2009, Yeh, Savin2017, Savin2020, DelHougne2020}.  
{We shall borrow this approach and transform it into the time domain} \cite{Xiao2016} {to account for system-specific early-time signals that are present in any particular realization of a complex scattering system.}
{This early-time RX signal rise-up comes from the non-statistical and system-specific prompt-response at the port location, and is present in many realizations of the cavity ensemble} \cite{Baranger1996, Brouwer1995, Richter2002, Hemmady2005, Muller2007, Yeh2011, Yeh2012}.
{Note that short orbits of open scattering systems should not be confused with closed periodic orbits, which are properties of closed billiard systems} \cite{Wintgen1987, Stein1992, Sieber2001}.

\begin{figure*}
\centering
\includegraphics[width=1\textwidth]{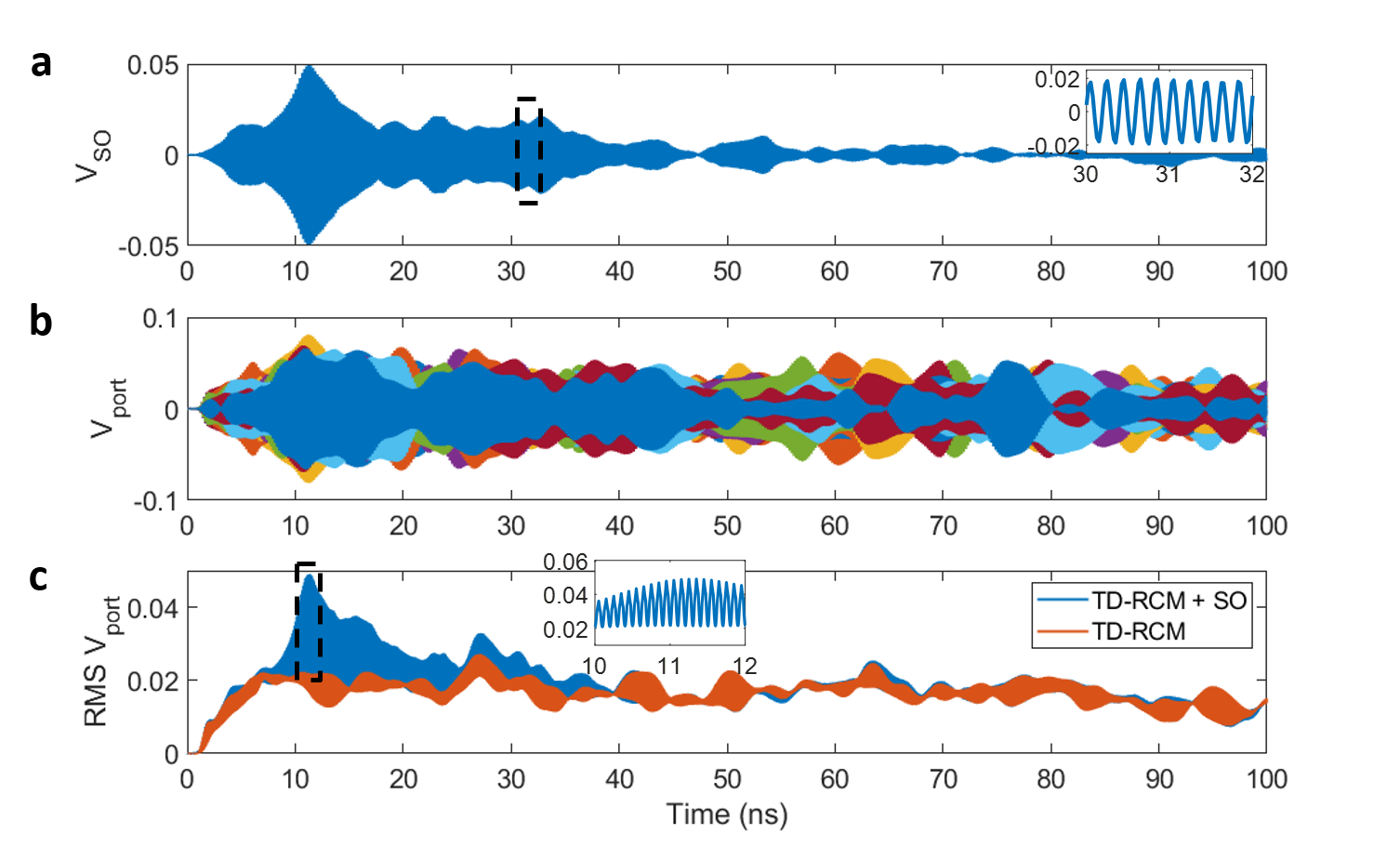}
\caption{\label{fig:addSO} 
\textbf{a.} shows the computed time-domain Short-Orbit (SO) signal at the gigabox RX port.
The injected pulse is a 5 GHz, 5 ns long flat single-frequency sine wave with a $\sim$ 1ns rise-time. The inset is the enlarged view of the $V_{SO}$ curve from 30 to 32 ns window (black dashed box).
The $V_{SO}$ is computed with FT methods using the injected pulse and the measured ensemble averaged S-parameter (see details in the main text).
\textbf{b.} consists of the TD-RCM simulated RX port signals $V_{port}$ from 20 realizations of the model system. Here the SO contribution $V_{SO}$ is added onto the TD-RCM computed $V_{port}$.
\textbf{c.} shows the RMS ensemble averaged RX port signal compiled from the port signals before and after the SO treatment. The inset is the enlarged view of the curve from 10 to 12 ns window (black dashed box).}
\end{figure*}

{To include short orbit effects in our time domain model,} Eqs. \ref{eq:modeV} to \ref{eq:knj}, {we add current-controlled voltage sources to each port voltage, $V_j \rightarrow V_j + V_{SO, \, j}$, where}
\begin{equation}\label{eq:16}
    V_{SO, \, j}(t) = \sum_{j'}(R_{rad, j} R_{rad, j'})^{1/2} \int_0^t dt'\hat{\zeta}_{jj'}(t - t') I_{j'}(t').
\end{equation}
{Here $\hat{\zeta}_{jj'}$ is a convolution kernel that describes the average transmission of wave amplitude from port $j$ to port $j'$.  
The kernel can be found by exciting one port and measuring the voltage at another.  
In the time domain, this would be done by imposing a step function current waveform at port $j'$ and measuring the voltage waveforms at all other ports over a period of time comparable to the length of the short orbits of interest.  These signals should then be multiplied by a decaying exponential to effectively perform a window average.  Alternatively, the ports $j'$ may be excited with monochromatic waves, and the response at other ports recorded for a range of injected frequencies covering the expected signal bandwidth.  
In this case, averaging would have to be conducted by changing the configuration of a stirrer or by window averaging in frequency.  
The frequency domain version of the short orbit contribution is given by }
\begin{equation}\label{eq:17}
    \Tilde{V}_{SO, \, j}(\omega) = \sum_{j'}(R_{rad, j} R_{rad, j'})^{1/2} \hat{\zeta}_{jj'}(\omega) \Tilde{I}_{j'}(\omega).
\end{equation}
{The inclusion of these terms}, Eq. \ref{eq:16} {in the time domain and} Eq. \ref{eq:17} {in the frequency domain, replicates the first term in the square bracket in} Eq. \ref{eq:tma4}.

{We note that if the receiving port voltages are measured with known injected current at the transmitting port, and with all receiving ports open-circuited, then} Eqs. \ref{eq:16} and \ref{eq:17} {can be used to find the appropriate transfer matrix elements $\hat{\zeta}_{jj'}(t)$ or ${\zeta}_{jj'}(\omega)$ directly. }
{If the receiving ports are not open-circuited, and a current flows in the receiving port, then this will cause the receiving port to re-radiate the signal, complicating the calculation of the various matrix elements.  
This generally will be a small effect when the diagonal impedance matrix elements are larger that the off-diagonal ones.  
As we have already assumed this we will neglect this reradiation effect and proceed. 
With $\hat{\zeta}_{jj'}(t)$ in hand we can then model nonlinear ports in time using} Eq. \ref{eq:16}.

Figure \ref{fig:addSO} (a) {shows an example of an averaged short orbit signal $V_{SO}(t)$ evaluated as follows.  
The frequency-dependent scattering matrix is measured by means of a microwave network analyzer for a $1$ $m^3$ enclosure containing a mode stirrer.  
Measurements are recorded for 50 positions of the mode stirrer and an average is taken.  
As the quantity that was actually measured was $S_{21}(\omega)$ with both ports connected to 50 Ohm transmission lines,  we seek an expression for the short orbit contribution either in the form of} Eq. \ref{eq:16} or \ref{eq:17}, {the relationship between these quantities must be determined.  
It is found to be}
\begin{equation}\label{eq:18}
    \langle S_{21} \rangle = \frac{2Z_0 (R_{rad, 1} R_{rad, 2})^{1/2}}{(Z_0 + Z_{rad, 1})(Z_0 + Z_{rad, 2})} \zeta_{21}(\omega).
\end{equation}
{Here we have assumed both transmitting and receiving ports are connected to identical transmission lines of impedance $Z_0$.  
We next assume a narrow band input signal $V_{in}(t)$ is incident on one of the ports.  
The assumed input signal is a 5 GHz sine wave, modulated by a 5 ns long flat top envelope with a 1 ns rise time. 
In practice, the input signal $V_{in}(t)$ is obtained by measuring the actual injected waveform (a 5 ns, 5 GHz sine wave).
The $V_{in}(t)$ is then Fourier transformed to the frequency domain and the average scattering coefficient is applied to arrive at an assumed frequency domain output signal.  
This signal is then Fourier transformed back to the time domain to arrive at a short orbit voltage signal ${V}_{SO}(t)$. 
This we plot in} Fig. \ref{fig:addSO} (a).  
{In the next section, we will be using this signal to compare measured and simulated signal properties.}

{We next combined the short orbit and statistical fluctuating contributions to the off-diagonal elements of the impedance matrix of} Eq. \ref{eq:tma4}.  
{This was done by generating an ensemble of 20 fluctuating matrices as given by different realizations of the random variables $\underline{c}$ in} Eq. \ref{eq:tma5}. 
{To complete the evaluations of} Eq. \ref{eq:tma5} {we estimated the average spacing between modes ($\Delta \omega_n$) and the Q factors of the modes in the range of 5 GHz} (see gigabox parameter values in Table \ref{tab:times}).  
{We then added $\xi_{jj'}$ to $\zeta_{jj'}$ in} Eq. \ref{eq:18} {and performed the same operations that produced $V_{SO}(t)$.  
These 20 signals are labeled $V_{port}(t)$ and are plotted in} Fig. \ref{fig:addSO} (b).  
Figure \ref{fig:addSO} {shows that the average and statistical contributions for the first 20 ns are comparable, but for long times the statistical signals are dominant.  
This comparison was taken further in} Fig. \ref{fig:addSO} (c) {where we extracted the time-dependent RMS envelope amplitude from the time-dependent signals resulting from 30 realizations of the fluctuating matrix with (blue) and without (red) the short orbit treatment and averaged them.  
In} Fig. \ref{fig:addSO} (b) {the presence of the short orbit signal is difficult to discern in the single realization signals.  
However, in} Fig. \ref{fig:addSO} (c) {there is a clear enhancement in the signal at around 12 ns that can be attributed to the short orbits.}

\section{III. Short-pulse experiment set-up}

\begin{figure*}
\centering
\includegraphics[width=0.95\textwidth]{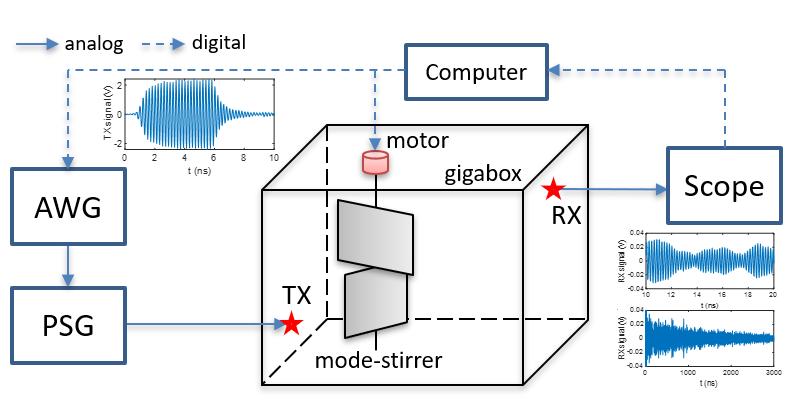}
\caption{\label{fig:fig0} 
Schematic of the short pulse time-domain experiment setup. 
The injected waveform (left inset, a 5 ns pulse) is generated using the arbitrary waveform generator (AWG) and the programmable signal generator (PSG).
The short pulse is broadcast into the enclosure through the transmitting (TX) port (red star).
The data ensemble is created by rotating a motorized mode-stirrer inside the enclosure.
The receiver (RX) port (red star) induced voltage signal is measured by the oscilloscope.
An example of the measured RX port signal is displayed on different time scales as the insets on the right.
A lab computer is used for instrument control and data transmission/collection.
}
\end{figure*}

To test the correctness and applicability of the TD-RCM method, we conducted short pulse injection experiments in the gigabox enclosure (shown schematically in Fig. \ref{fig:fig0}).
The gigabox is a metallic enclosure with a volume of $\sim 1$ $m^3$.
The round-trip time for short ray trajectories in the gigabox is $T_{rt} = 7$ ns.
A mode-stirrer, consisting of two thin metallic panels, is located inside the gigabox to randomize the EM fields.
One can rotate the angle of the mode-stirrer through many fixed values to obtain different scattering details, thus creating an ensemble of data.
To automate the measurement process, we installed a motor outside the gigabox to control the rotation angle of the mode stirrer.
Note that we conduct 2-port pulse injection experiments.
One of the locations serves as the input port, where a single carrier frequency $f_c$ is modulated by a 20 ns short pulse envelope and is injected into the gigabox.
To generate the input pulse, we use the Tektronix AWG7052 5 GS/s arbitrary waveform generator (AWG) to produce a square pulse and then use the Agilent programmable signal generator (PSG) E8267D to act as a modulated waveform source with a fixed center frequency $f_c$.
We use a series of different $f_c$ values ranging from $5$ to $10$ GHz in the experiment. 
A large number of system modes will be excited by these pulses.
For example, there exists about $2 \times 10^4$ modes from $4.5 \sim 5.5$ GHz, and about $9 \times 10^4$ modes from $9.5 \sim 10.5$ GHz inside the gigabox.
The pulses are injected into an initially quiet cavity. 
A non-zero induced voltage signal will be measured at the RX port.
A 50 $\Omega$ load is connected to the RX port.
Typical input and output waveforms are shown as insets in Fig. \ref{fig:fig0} where one should note the difference in time scales between the TX and RX waveforms.
We rotate the mode-stirrer to 200 fixed locations for each choice of the center frequency to create an ensemble of systems and measure the resulting received time-domain waveforms at the RX port using an Agilent DSO91304A 40 GS/s oscilloscope.
As illustrated by the insets of Fig. \ref{fig:fig0}, we capture the detailed time-domain signal evolution on the sub-period time scale, allowing us to examine details of the induced voltage such as extreme values and to compile complete histograms of the time evolution of the induced voltages.
Statistical analysis is then done on this ensemble of time-domain waveforms.
{We have summarized the characteristic time scales of the gigabox enclosure in Table} \ref{tab:times}.

\begin{table*}
\begin{ruledtabular}
\begin{tabular}{l l l}
Characteristic Time Scales & Gigabox Parameter Values & Descriptions\\
  \hline
$T_{osc} = 1/f_c$ & $0.2 ns$ ($f_c = 5GHz$) & Input carrier oscillation period \\
$T_{rt} = 2(V^{1/3})/c$ & 7 ns ($V = 1 m^3$) & Round-trip time\\
$T_{BU} = Q/\omega_c$ & $1.5\mu s$ (Q = 4.7e4 at 5GHz) & Build-up time\\
$T_{H} = 2\pi/\Delta \omega_n$ & $23.3\mu s$ (evaluated at 5GHz) & Heisenberg time\\
\end{tabular}
\end{ruledtabular}
\caption{\label{tab:times} Summary table of different system time scales for the Gigabox enclosure used in the experiments described here.
The mean-mode spacing $\Delta \omega_n$ is defined above in section I.}
\end{table*}

\section{IV. Late-time Statistical Tests}

As introduced in the previous section, we have conducted short-pulse injection experiments in an electrically-large ray-chaotic cavity to test the performance of the TD-RCM model.
Here we study the late-time (beginning at $7T_{rt} \sim 50$ ns after the pulse injection) quantities where the short orbit effects are not present and a reverberant field structure has been established.
We simulate the experimental structure using realistic information for the experimental setup with the TD-RCM method.
More specifically, the information required are the gigabox loss parameter $\alpha$ \cite{Gradoni2014} at frequency $f_c$, the injected waveform $V_{in}(t)$, the mean mode-spacing, the antenna radiation resistance $R_{rad}$ over the bandwidth of the pulse for each port, and the value of the linear load impedance.
The gigabox TD-RCM simulation is run 1000 times for each center frequency, where new lists of system modes $\omega_n$ and all random port-mode coupling coefficients $c_{jn}$ are refreshed each time.

\begin{figure}
\centering
\includegraphics[width=0.5\textwidth]{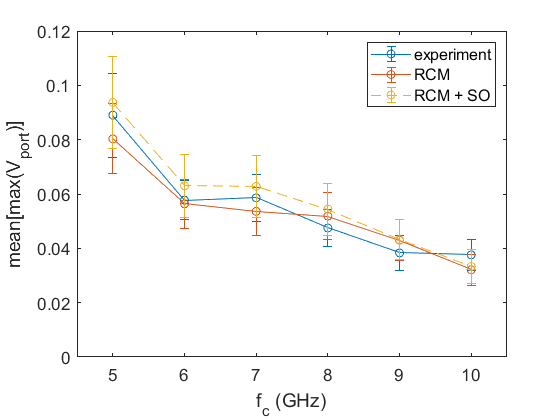}
\caption{\label{fig:tdfig2} 
Late-time extreme port-voltage value statistics.
We show the mean values, and their standard deviations, of the receiver port maximum voltage data over 200 realizations as a function of the center frequency of the input pulse. 
The data measured experimentally are shown in solid blue, and the data computed by TD-RCM and the TD-RCM with the SO addition are shown in solid red and dashed yellow, respectively.
The input pulse is a single-frequency 20 ns modulated sine pulse. The center frequency $f_c$ is varied from 5 GHz to 10 GHz.
Note that the input pulse has a finite rise/fall time ($1 \sim 2$ ns, see Fig. \ref{fig:fig0} left inset).
The error bar is one standard deviation from the mean over the ensemble.
}
\end{figure}

We first study the statistics of the maximum receiver (RX) port voltage (max[$V_{port}$]) in the time domain. 
For each configuration in the ensemble (typically 200 different system configurations generated in the experiment by rotating the stirrer, and generated in the TD-RCM by generating a new random vector of coupling coefficients $c_{jn}$ and mode frequency $\omega_n$ that appear in Eqs. (\ref{eq:modeV}) and (\ref{eq:portV}) we record the maximum receiver port voltage over time. 
We then average these maxima over the members of the ensemble to arrive at a quantity mean[max($V_{port}$)]. 
We observe the value of this quantity to be stable once the ensemble size is over 50.  
This process is then repeated for different carrier frequencies.
We study the RCM simulations with or without the SO addition treatment.
We compare the statistics of max$\left[ V_{port} \right]$ obtained from the experiment and the TD-RCM simulations.
As summarized in Fig. \ref{fig:tdfig2}, we have found a relatively good agreement for mean[max$(V_{port})$] between the TD-RCM model prediction (red and yellow) and experimental data (blue).
All data points fall within one standard deviation of each other.
Six different experiments are considered here, where we varied the center frequency of the injected sine wave from 5 GHz to 10 GHz in steps of 1 GHz.
Note that the value of the RX port signal decreases as the center frequency increases.
This can be explained by the loss parameter of the gigabox, which increases from $\alpha \sim 3$ (5 GHz) to $\alpha \sim 12$ (10 GHz).
Note that, as shown in Fig. \ref{fig:tdfig2}, the data computed from both the RCM model and the RCM model with the SO addition show relatively good agreement with the experimental data.
Two observations can be made.
First, the SO correction does not significantly alter the long-term statistics of the original TD-RCM predictions.
Second, the maximum value of the RX port signal is a quantity that is not strongly affected by the early-time SO contributions.
This is a consequence of the high value of the quality factor for the experimental cavity.
The build-up time of the cavity modes is longer than the round-trip time $T_{rt}$ for short orbits (see Table I).

\section{V. Early-time Statistical Tests}

\begin{figure*}
\centering
\includegraphics[width=1\textwidth]{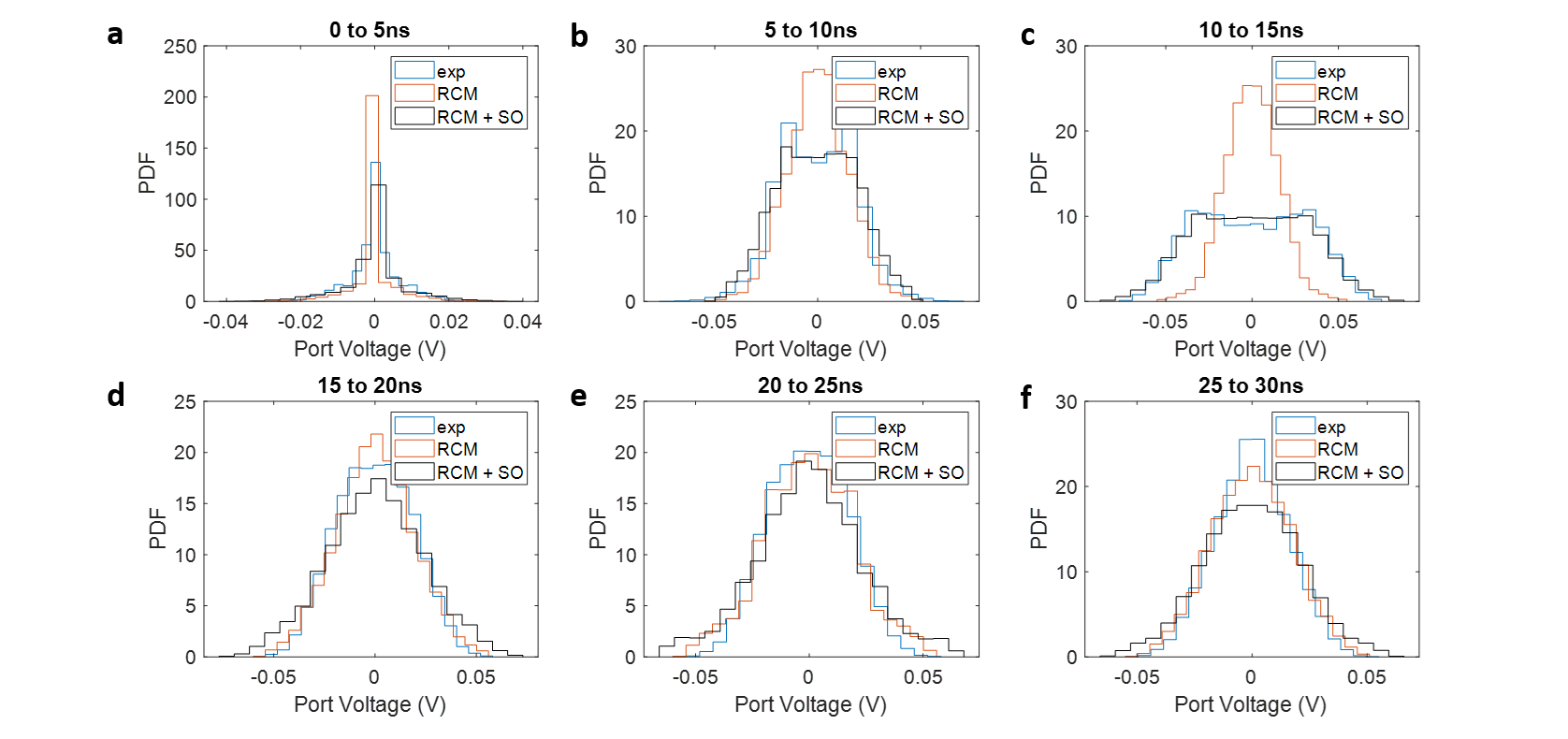}
\caption{\label{fig:VrxPDF} 
Comparison of RX received port voltage statistics in different time intervals for the gigabox experiment and related simulations.
The statistics of the distribution of the RX port voltages are shown from 3 data sets: the experimental data, the TD-RCM simulation neglecting SO, and the TD-RCM simulation with the SO addition.
A 5 GHz, 5 ns modulated single frequency sine pulse is adopted as the injected waveform into the gigabox for all three tests.
We show the RX port voltage statistics in a series of 5 ns windows. The window start time is swept from 0 to 25 ns in 5 ns steps, shown in (a) to (f), respectively.}
\end{figure*}

In Section II. C we have introduced a method to include the direct short-orbit (SO) contribution to the TD-RCM.
The proposed method computes the time-domain SO contribution from a TX port to a RX port ($V_{SO}$), and directly adds $V_{SO}$ to the TD-RCM computed RX port voltages ($V_{port} \rightarrow V_{port} + V_{SO}$).
Here we study the effectiveness of the SO addition method by studying the detailed distribution of $V_{port}$ in the time domain.
We focus on the early-time ($< 5\, T_{rt}$ after the pulse injection) dynamics when the SO contribution dominates the induced voltage at the RX port.
We compare $V_{port}$ histogram data obtained from three different tests: the gigabox short-pulse experiments (see details in Section III), the corresponding TD-RCM simulations that neglect SO, and the TD-RCM simulation that includes the SO contribution.
The distributions of the $V_{port}(t)$ values for different time intervals are shown in Fig. \ref{fig:VrxPDF}.
Because the injected pulse duration is 5 ns, we set the time window as 5 ns and look at the relatively earlier times (before $5T_{rt}$) in $V_{port}(t)$.
The experimental and the RCM + SO histograms consistently show good agreement throughout the time evolution of the waveform.
However, the RCM-only results show a clear deviation from the experimental results during the 5 to 15 ns (1-3 times $T_{rt}$) time interval.
We believe the RCM-only results show smaller values of $V_{port}(t)$ due to a lack of SO-delivered energy.
As shown in Figs. \ref{fig:addSO} (a) and (c), the time-domain SO contribution shows the strongest amplitude contribution from 10 to 15 ns.
Inside the same time window, we also observe the most significant deviation between the RCM-only $V_{port}$ statistics and the other two data sets (Fig. \ref{fig:VrxPDF} (c)).
At a later time (15 ns after the pulse injection), we find that the $V_{port}$ statistics from the three data sets show a relatively good agreement (Figs. \ref{fig:VrxPDF} (d-f)).
We find that the late-time $V_{port}$ distribution fits nicely with a Gaussian distribution (not shown here) \cite{Hill1998}.
On the other hand, a substantial deviation from Gaussian distribution in $V_{port}$ statistics marks the existence of substantial SO contributions in the system.
Examples of strong SO effects are the experimental and the RCM + SO datasets in Figs. \ref{fig:VrxPDF} (b) and (c).

\section{VI. Nonlinear TD-RCM Application: Reverberant Wave Reservoir Computing Hardware}

Here we aim to demonstrate one successful application of the TD-RCM method in a problem that requires a nonlinear port treatment in the time-domain.
Recently in Ref. \cite{Ma2022}, we have experimentally realized a physical system, namely reverberant wave reservoir computing (RC) hardware, which can execute multiple machine learning tasks on time-domain signals.
{Reservoir computing (RC) is a type of machine-learning algorithm that utilizes a complex system with a large number of nonlinear computing nodes} \cite{Pathak2018}.
{In a typical RC structure, the input signal is randomly mapped to a high-dimensional `reservoir layer', which consists of connected nonlinear nodes.
The system output is extracted out of the reservoir layer through a linear transfer matrix, $W_{out}$.
The training of an RC is only conducted to the optimization of the $W_{out}$ matrix} \cite{Lu2016, Pathak2018}.
{An RC realization requires a system with a complex and diverse set of modes, and requires a nonlinear evolution of the system state. }

The RC hardware is based on a wave chaotic enclosure with multiple diode-loaded ports.
To operate the wave-based RC, one injects the input data stream into a wave chaotic enclosure through a linear port, and then measures the induced voltage signal at nonlinear ports at several discrete locations.
{The nonlinear induced voltage signals at the ports will in turn excite cavity modes at harmonic frequencies.
In this manner, one introduces a nonlinearity and further complexity into the modal dynamics of the cavity RC.}
These voltage signals effectively serve as the `neuron' responses in the traditional software RC.
After the training, the desired output signal is obtained by properly summing up the RC neuron responses.
{A ray-chaotic microwave cavity with diode-loaded output ports is very well suited for this task.  
We use TD-RCM with nonlinear ports to simulate the microwave realization of an RC.
The practical resonator system to be simulated with TD-RCM is a quasi-2D resonator.
The shape of the billiard is similar to a quarter of a bow tie with an area of $A=0.115 m^2$.
The characteristic length of the billiard is $A^{0.5} \sim 0.35m$.
In the later TD-RCM simulation, the input waveform is set to center at $f_c=4GHz$.
With a height of $d=7.9mm$, the billiard is considered to be quasi-2D because the electric field is polarized in the z-direction for all frequencies $f<c/(2d)=18.9GHz$} \cite{So1995}.

\begin{figure*}
\centering
\includegraphics[width=1\textwidth]{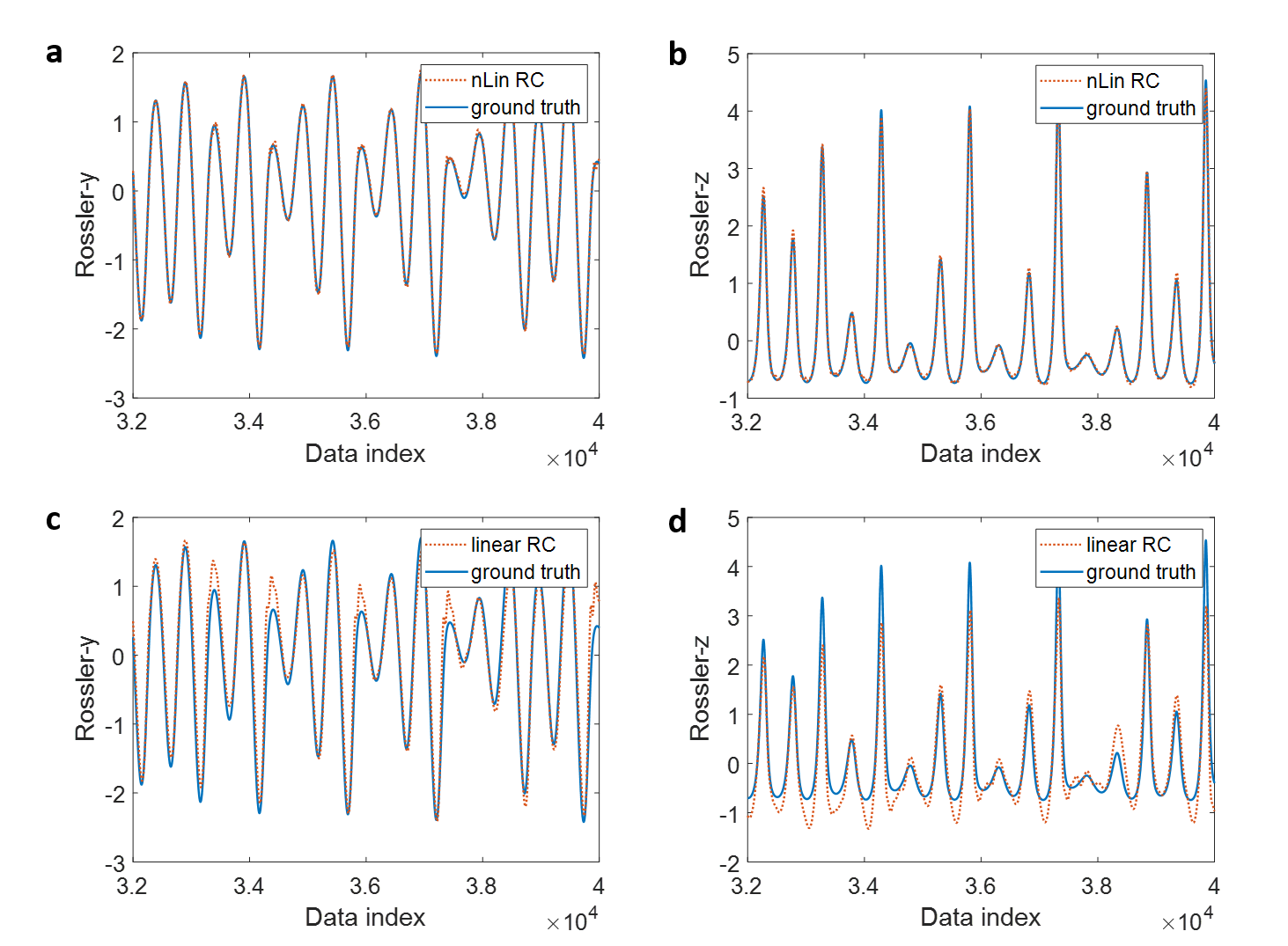}
\caption{\label{fig:Rossler} 
The Rossler-observer task testing set performance.
The trained RC is given a continuous stream of the x(t) data generated from the coupled nonlinear Rossler equations and has to deduce the evolution of the y(t) and z(t) time series based only on training data.
(a) and (b) show the Rossler-y and Rossler-z series results produced by the nonlinear RC, while (c) and (d) show the Rossler-y and Rossler-z series results produced by a linear RC, respectively.
The TD-RCM simulated reservoir computing data are shown in red dotted lines. The actual states are shown in solid blue lines.
The x-axis shows the index of the testing, ranging from the $32000^{th}$ to the $40000^{th}$ data points.
Each data index corresponds to a time bin of duration 250 ps.
}
\end{figure*}

Here we utilized the nonlinear version of TD-RCM to simulate the dynamics of the wave-based RC hardware.
The diode-loaded ports are modeled as in Section II.B.
We have put a single linear input port and 100 diode-loaded output ports in our TD-RCM cavity model computation.
All diodes have the same characteristics ($R = 50\Omega, C = 1.5e^{-12}F, I_0 = 1e^{-7}A$, and $T = 300K$).
The specific type of machine learning (ML) task we studied here is the so-called Rossler observer task \cite{Lu2016}.
The Rossler system is governed by these equations,
\begin{equation} \label{eq:Rossler}
\begin{aligned}
    \Dot{x} &= -y-z,\\
    \Dot{y} &= x+ay,\\
    \Dot{z} &= b+z(x-c),\\
    \end{aligned}
\end{equation}
where $a=0.5, b=2.1, c=3.5$, and the over-dot denotes derivative with respect to time.
In this task, the ML model is expected to predict the Rossler attractor $y$- and $z$-component data when only the $x$-component information is supplied.
More details of the Rossler observer task setup can be found in Refs. \cite{Lu2016, Ma2022}.
The wave-based RC training is completed with the out-coupling matrix $W_{out}$, which is a many-to-few linear mapping of the reservoir state (as measured through the 100 diode-loaded ports) to $X$ output channels ($X$ being the dimension of the final output state).
The process of finding the optimized $W_{out}$ matrix is completed by running the Ridge regression algorithm on a digital computer \cite{Lu2016}.
Here, the training phase used $80\%$ of the entire dataset (80000 data points uniformly sampled in $\sim 200$ oscillation periods).

With optimized cavity parameters (with loss parameter $\alpha = 300$ and a 35 V input amplitude for the Rossler $x(t)$ signal), the testing set results are shown in Fig. \ref{fig:Rossler}.
In this case, the trained RC is fed the $x(t)$ time domain waveform generated by an arbitrary waveform generator with a time step size of 250 ps.  
The RC has not seen this waveform in its training phase.
The task for the RC is to continuously generate the $y(t)$ and $z(t)$ time domain signals without knowledge of the underlying equations of motion for the Rossler attractor.
The performance of the RC prediction can be evaluated using the normalized mean square error, e.g., defined for the $y$-variable as NMSE $= \sum_n[s_n - s_n']^2/\sum_n s_n^2$, where $s'$ and $s$ denote true and RC-inferred values for the Rossler time series, respectively, and $n$ is the time index of the testing set data.
The $s'$ true signal is denoted as ``ground truth'', and the RC-inferred signal $s$ is denoted ``nLin RC'' in Fig. \ref{fig:Rossler} (a) and (b).
The NMSE is computed to be 0.038 and 0.076 for the results shown in Fig. \ref{fig:Rossler} (a) and (b), respectively, showing that the TD-RCM simulated RC can perform this ML task with high accuracy.
We note that NMSE generally decreases with an increasing number of output ports.
In Fig. \ref{fig:Rossler} (c) and (d) we show the deduced Rossler series from a linear RC system (no diode connected to the ports).
The NMSE is computed to be 0.228 and 0.34 for the results shown in Fig. \ref{fig:Rossler} (c) and (d), respectively, 
It is clear that the existence of nonlinearity is crucial to the success of a reservoir computer system.

With TD-RCM, one can study the dynamics of the reverberant wave RC with great fidelity and detail because all system parameters can be easily controlled.
It is also possible to study the cavity realization of the RC with electromagnetic numerical simulation software, such as CST \cite{Ma2022}. 
However, the computing speed of TD-RCM is much faster than the numerical EM simulation due to the latter's need for an accurate CAD drawing and the corresponding need for detailed spatial-domain mesh-solving.
{For comparison purposes, we also conducted time-domain simulations of the Rossler-observer task experiment with CST. 
The CST simulation model has diodes connected to the ports and also has an accurate 3D resonator structure.
The TD-RCM and the CST studies have the same injected signal, number of diode-loaded ports, and system Q-factor.
The CST simulation takes roughly 66s to simulate 1ns ahead of time while the TD-RCM codes take roughly 23s. 
Note that the advantage of TD-RCM's speed enhancement is lower than the linear case (discussed in the next section) because TD-RCM takes extra time when solving the transcendental equation }(\ref{eq:diode3}).

\section{VII. Discussion and Summary}

The TD-RCM is a statistical treatment of time-dependent induced voltage signals in electrically-large complex enclosures.  
It eschews geometrical details and utilizes the random plane wave hypothesis to describe the statistical properties of the cavity modes.
A significant limitation of TD-RCM is the neglect of the nonzero time for signals to propagate through the cavity. 
In making the random plane wave approximation, the model neglects the correlation between modes that carry the early-time information about the location of the ports.
The inclusion of the short orbit time domain response restores, to a degree, this information.
{The causality of computed time-domain signals in many frequency-domain-based methods faces the issue of nonzero signals at negative times.
Such causal concern is absent in TD-RCM as the amplitude of all system mode/port signals will stay at zero unless an external excitation pulse is injected.
On the other hand, the model implicitly assumes that an excited mode develops and varies its amplitude simultaneously everywhere in the enclosed space of the cavity. 
Thus the ports are instantaneously excited by all the modes, giving rise to a coupling between ports that is not causal.}
{One will encounter causality concerns for the cases where the pulse length is not substantially longer than (i.e., 10 times larger) the transit time for the pulse across the system.
To address this issue, one can apply a delay to all TD-RCM computed time-domain results to account for a longer transit time.}

Another practical limitation of TD-RCM is the trade-off between computing speed and system mode density.
The TD-RCM models each and every one of the $N$ system eigenmodes that are excited by an input pulse. 
We can estimate the required number of eigenmodes $N=BW/(\Delta f|_{f_c})$, where $BW$ is the bandwidth of the injected pulse, and $\Delta f|_{f_c}$ represents the system mean-mode-spacing at the pulse center frequency $f_c$.
Here is an estimate of the scale of $N$ in our gigabox short-pulse experiment.
For a 5 ns pulse at $f_c = 5$ GHz, there are $N \sim 8500$ modes excited.
Hence, the computation speed of the TD-RCM model is limited by the total number of modes that must be modeled.
A method of systematically reducing the total number of degrees of freedom to simulate is desired.
{As a reference, we list the computation resources of one run of the TD-RCM simulation conducted in Fig}. \ref{fig:addSO} {.
The simulation was conducted with a desktop computer that has 11th Gen Intel(R) Core(TM) i7.
For a simulation with 1 input port, 1 output port, and 20000 system modes, the simulation takes 0.5s to simulate the dynamics of a 1 ns signal when utilizing 50\% of the CPU resources.
The maximum RAM consumption was around 7GB during the computation.
We note that the RAM consumption is related to the total number of system modes.
As a head-to-head comparison, we have set up a time-domain simulation in CST.
The simulation model has the same 3D size as the experimental structure but with a simplified interior geometry (no mode stirrer).
The CST simulation takes roughly 100s to simulate the dynamics of a 1ns signal, and the simulation utilizes the full capacity of the CPU.
The TD-RCM simulation presents a 200 times improvement in computation speed, and the computation resources it consumed may also be optimized by using different coding environments.}

The workflow of TD-RCM starts by generating a list of cavity modes using RMT.
Each of the cavity modes is characterized by a driven damped harmonic oscillator equation.
A random summation computes the port signal over all excited cavity modes.
A TD-RCM user will need to input the basic enclosure information (RCM loss parameter $\alpha$, or equivalently Q-values), port information ($R_{rad}$, port load impedances, and voltage-current relationship for all nonlinear port loads), and the injected waveform.
An ensemble can be created by changing the list of eigenmodes and the mode-port random coupling coefficients.
One can then study many statistical properties of the time-dependent port voltages and system-mode dynamics, only a few of which have been discussed here.

{In this manuscript, we proposed the TD-RCM model which can simulate the temporal evolution of both port and cavity mode signals in chaotic systems.
To validate the TD-RCM, we have conducted short-pulse injection experiments into a large complex enclosure and performed statistical analysis of the measured port voltage signal $V_{port}(t)$.
We have found good agreement between the measured and the TD-RCM simulated port voltage signals in two types of statistical analysis, which are the statistics of the peculiar/rare incidents (maximum receiver (RX) port voltage) in the time domain, and the overall port voltage value distribution. 
Besides successfully simulating the effects of background modes, we have included the effect of short orbits.
The advantage of the TD-RCM to simulate the time-domain response of nonlinear chaotic systems is demonstrated in section VI.}
We have also worked out the formalism for including nonlinear port loads and demonstrated its use in Appendix C.

Concerning applications, one can use TD-RCM to simulate both the linear and nonlinear time-reversed wave focusing process between two ports in a complex scattering environment \cite{lerosey2004time, lerosey2006time, anlage2007new, taddese2010sensing, Frazier2013, hong2013focusing, Frazier2013a, hong2014nonlinear, Hong2015, Xiao2016, Frisco2019, 9130060}.
TD-RCM can also be applied to coda wave interferometry applications \cite{snieder2002coda, lobkis2003coda, Kuhl2005, snieder2006theory}, and the analysis of scattering fidelity studies \cite{schafer2005fidelity, gorin2006dynamics, taddese2009sensor}.
The reverberant wave-based reservoir computing hardware can also be studied with the nonlinear TD-RCM treatment.
A TD-RCM user can set up a collection of SO signals to study the coherent energy delivery between the ports when the cavity geometry is known \cite{Xiao2016}.
For future directions, one may study the nonlinear properties of diode-loaded port with both TD-RCM and experimental methods \cite{Ma2020,Ma2020b}.
Another future direction is the inclusion of large multi-mode apertures between complex enclosures in TD-RCM.
Finally, TD-RCM may also be applied to simulate the linear and nonlinear dynamics for other systems which RMT can model.
Examples range from optical pulses in multi-mode fibers \cite{Doya2002, Xiong2016} to financial market fluctuations \cite{Fenn2011}.

\section{Acknowledgements}

We acknowledge discussions with Bisrat Addissie and Zachary Drikas of the US Naval Research Laboratory.  
This work was supported by ONR under Grant No. N000141912481, DARPA Grant HR00112120021, and the Maryland Quantum Materials Center.

\section{Appendix}

\subsection{A. Generation of System Eigenmodes with RMT}

One will need to provide a list of eigenmodes of the closed system to initiate TD-RCM.
It is possible to obtain all cavity eigenmodes accurately using numerical simulation tools (CST, HFSS, COMSOL).
However, such a process would be time-consuming for an electrically large system and impossible when the accurate geometrical details are unknown.
Here we utilize the property of a chaotic system to simulate a list of system modes by only knowing the macroscopic system information (i.e., volume, area) and the operating frequency \cite{Gradoni2014}.

As introduced in the main text, a list of system eigenmodes $\omega_n$ can be generated using Random Matrix Theory (RMT). 
We start by computing the eigenvalues $\lambda_{rmt}$ of large random matrices.
The matrix size equals the total number of system modes that we want to model, and the matrix elements are drawn from a Gaussian distribution controlled by the symmetry group of the system \cite{Dyson1962}.
The main symmetry classes of concern are the Gaussian Orthogonal and Gaussian Unitary ensembles of random matrices.
The system eigenmodes $\omega_n$ are computed from the relation $\lambda_{rmt} = \frac{\omega_c^2 - \omega_n^2}{\Delta \omega^2}$, where the quantity $\omega_c$ is the center frequency of the band that we want to model, and $\Delta \omega^2$ is the mean-mode-spacing near $\omega_c$.
To sum up, one first uses the total number of modes ($N$) to generate $N$ random matrix eigenvalues $\lambda_{rmt}$, and then uses the center frequency and mean-mode-spacing to compute $N$ system modes $\omega_n$ from $\lambda_{rmt}$.

\subsection{B. Derivation of Time Domain Random Coupling Model}

In this section, we outline the derivation of TD-RCM.
We start by representing the EM field inside the closed cavity in a basis of modes, $\mathbf{e}_n, \mathbf{h}_n, \phi_{n'}$, where
\begin{equation} \label{eq:eq1}
\begin{aligned}
    & \nabla \times \mathbf{e}_n = k_n \mathbf{h}_n\\
    & \nabla \times \mathbf{h}_n = k_n \mathbf{e}_n\\
    & \nabla^2 \phi_{n'} + \beta_n^2 \phi_{n'} = 0 \\
\end{aligned} \quad.
\end{equation}
The quantities $\mathbf{e}_n$ and $\mathbf{h}_n$ are the electric and magnetic fields for the $n^{th}$ cavity mode, $\phi_{n'}$ is potential of the electrostatic modes, and $\beta$ and $k$ are the corresponding wavenumbers.
For boundary conditions, we have the tangential components of $\mathbf{e}_n = 0$ at the boundary and the normal component of $\mathbf{h}_n = 0$. 
For electrostatic modes, we have $\phi_{n'} = 0$ on the boundary.
Between two different cavity modes (mode $m$ and mode $n$, $m \neq n$), we have the orthogonal relationship
\begin{equation} \label{eq:ortho}
\begin{aligned}
    & \int_V d^3x \, \mathbf{e}_m \cdot \mathbf{e}_n = 0 \\
    & \int_V d^3x \, \mathbf{h}_m \cdot \mathbf{h}_n = 0 \\
\end{aligned} \quad.
\end{equation}
The quantity $V$ is the cavity volume.
We define a normalization process,
\begin{equation} \label{eq:normal}
    \int_V d^3x \, \mathbf{e}_n \cdot \mathbf{e}_n = k_n^{-1} = \int_V d^3x \, \mathbf{h}_n \cdot \mathbf{h}_n. \\
\end{equation}
Similarly, for the electrostatic modes, we have
\begin{equation} \label{eq:ortho2}
\begin{aligned}
    & \int_V d^3x \, \nabla \phi_{m'} \cdot \phi_{n'} = 0, \quad m' \neq n', \\
    & \int_V d^3x \, \nabla \phi_{n'} \cdot \phi_{n'} = \beta_{n'}^{-1}. \\
\end{aligned}
\end{equation}

We next expand the total EM field inside the cavity $\mathbf{E}$ and $\mathbf{H}$ in the basis of cavity modes, 
\begin{equation} \label{eq:modes}
\begin{aligned}
    & \mathbf{E} = \sum_n U_n \mathbf{e}_n + \sum_{n'} U_{n'} \nabla \phi_{n'} \\
    & \mathbf{H} = \sum_n I_n \mathbf{h}_n \\
\end{aligned} \quad.
\end{equation}
Here we use $U_n$ and $I_n$ to represent the voltage and current of mode $n$.
Then we plug Eq. \ref{eq:modes} to Maxwell's equations.
We dot Faraday's law with $\mathbf{h}_n$ and integrate over volume,
\begin{widetext}
    \begin{equation} \label{eq:max1}
        -\mu \frac{\partial}{\partial t} \int_V d^3x \, \mathbf{h}_n \cdot \mathbf{H} = -\frac{\mu}{k_n} \frac{\partial}{\partial t} I_n = \int_V d^3x \, \mathbf{h}_n \cdot \nabla \times \mathbf{E} = \int_V d^3x \, \mathbf{E} \cdot \nabla \times \mathbf{h}_n = U_n.
    \end{equation}
\end{widetext}
We dot Ampere's law with $\mathbf{e}_n$, 
\begin{widetext}
    \begin{equation} \label{eq:max2}
        \epsilon \frac{\partial}{\partial t} \int_V d^3x \, \mathbf{e}_n \cdot \mathbf{E} = \frac{\epsilon}{k_n} \frac{\partial}{\partial t} U_n = -\int_V d^3x \, \mathbf{e}_n \cdot \mathbf{J} + k_n\int_V d^3x \, \mathbf{h}_n \cdot \mathbf{H}.
    \end{equation}
\end{widetext}

The quantities $\epsilon$ and $\mu$ are the permittivity and permeability of the system environment.
The current density $\mathbf{J}$ above has two components: the leakage current due to bulk conductivity representing cavity losses, and the driving current due to port current $I_j$ ($j$ is the index of the port),
\begin{equation} \label{eq:current}
    \mathbf{J} = \sigma \mathbf{E} + \sum_{ports} I_j \mathbf{u}_j.
\end{equation}
Here $\mathbf{u}_j$ is a profile function for the port $j$ current density. 
It is normalized with unit magnitude has units $L^{-2}$.
The quantity $\sigma$ is the conductivity of the cavity wall.
Then the first term in Eq. \ref{eq:max2} becomes
\begin{widetext}
    \begin{equation} \label{eq:e7}
        \int_V d^3x \, \mathbf{e}_n \cdot \mathbf{J} = \int_V d^3x \, \mathbf{e}_n \cdot \left( \sigma \mathbf{E} + \sum_{ports} I_j \mathbf{u}_j \right) = \frac{\sigma}{k_n} U_n + \sum_{ports} I_j \hat{c}_{nj}.
    \end{equation}
\end{widetext}
Here the dimensionless coupling coefficient is defined $\hat{c}_{nj} = \int_V d^3x \, \mathbf{e}_n \cdot \mathbf{u}_j$, and the Ampere's law (Eq. \ref{eq:max2}) becomes
\begin{equation} \label{eq:amp2}
    \frac{\epsilon}{k_n} \frac{\partial}{\partial t} U_n + \frac{\sigma}{k_n} U_n = - \sum_{ports} I_j \hat{c}_{nj} + I_n.
\end{equation}
The port current $I_j$ will also excite an electrostatic field component,
\begin{equation} \label{eq:amp3}
    \epsilon \frac{\partial}{\partial t} \int_V d^3x \, \nabla \phi_{n'} \cdot \mathbf{E} = \frac{\epsilon}{\beta_{n'}} \frac{\partial}{\partial t} U_{n'} = - \frac{\sigma}{\beta_{n'}} U_{n'} - \sum_{ports} V_j \hat{c}'_{n'j}.
\end{equation}
Similarly we introduce a coupling coefficient $\hat{c}'_{n'j} = \int_V d^3x \, \nabla \phi_{n'} \cdot \mathbf{u}_j$.
Finally, we have the induced electric field at the port $j$ written as
\begin{equation} \label{eq:portVV}
    V_j = - \int_V d^3x \, \mathbf{u}_j \cdot \mathbf{E} = - \sum_{em-modes} U_n \hat{c}_{nj} - \sum_{es-modes} U_{n'} \hat{c}'_{n'j}.
\end{equation}

We write the complete time domain equations here
\begin{equation} \label{eq:all1}
\begin{aligned}
    -\mu \frac{\partial}{\partial t} I_n &= k_n U_n \\
    \frac{\epsilon}{k_n} \frac{\partial}{\partial t} U_n + \frac{\sigma}{k_n} U_n &= - \sum_{ports} \hat{c}_{nj} I_j + I_n \\
    \frac{\epsilon}{\beta_{n'}} \frac{\partial}{\partial t} U_{n'} + \frac{\sigma}{\beta_{n'}} U_{n'} &= - \sum_{ports} \hat{c}'_{n'j} I_j \\
\end{aligned} \quad.
\end{equation}
The coupling coefficients are $\hat{c}_{nj} = \int_V d^3x \, \mathbf{e}_n \cdot \mathbf{u}_j$ and $\hat{c}'_{n'j} = \int_V d^3x \, \nabla \phi_{n'} \cdot \mathbf{u}_j$. 
We rewrite Eq. \ref{eq:all1} by eliminating the mode currents ($I_n$),
\begin{equation} \label{eq:all2}
\begin{aligned}
    & \frac{\partial^2}{\partial t^2} U_n + \frac{\sigma}{\epsilon} \frac{\partial}{\partial t} U_n + \frac{k_n^2}{\epsilon\mu} U_n = - \sqrt{\frac{\mu}{\epsilon}} \frac{k_n}{\sqrt{\mu \epsilon}} \sum_{ports} \hat{c}_{nj} \frac{\partial}{\partial t} I_j\\
    & \frac{\partial}{\partial t} U_{n'} + \frac{\sigma}{\epsilon} U_{n'} = - \frac{\beta_{n'}}{\epsilon} \sum_{ports} \hat{c}'_{n'j} I_j. \\
    & V_j = - \sum_{em-modes} U_n \hat{c}_{nj} - \sum_{es-modes} U_{n'} \hat{c}'_{n'j}.
\end{aligned}
\end{equation}

Here we study the port-mode coefficients $\hat{c}_{nj}$ in Eq. \ref{eq:all2} by switching to the frequency domain version.
We have 
\begin{equation} \label{eq:f1}
    (\omega_n^2 - i\omega \gamma - \omega^2) U_n = i \sqrt{\frac{\mu}{\epsilon}} \omega_n \omega \sum_{ports} \hat{c}_{nj} I_j
\end{equation}
where $\gamma = \frac{\sigma}{\epsilon}$ and $\frac{k_n}{\sqrt{\mu \epsilon}} = \omega_n$.
The electrostatic modes are characterized by $(\gamma - i\omega) U_{n'} = - \frac{\beta_{n'}}{\epsilon} \sum_{ports} \hat{c}_{n'j} I_{n'}$.
The port voltage is written as
\begin{widetext}
    \begin{equation} \label{eq:f2}
        V_j = -i \omega \sqrt{\frac{\mu}{\epsilon}} \sum_{em-modes} \sum_{ports} \frac{\omega_n \hat{c}_{nj'} \hat{c}_{nj}}{(\omega_n^2 - i\omega \gamma - \omega^2)} I_{j'} + \sum_{es-modes} \sum_{ports} \frac{\beta_{n'}}{\epsilon} \frac{\hat{c}_{n'j'} \hat{c}_{n'j}}{\gamma - i \omega} I_{j'}
    \end{equation}
\end{widetext}

where $j$ and $j'$ are the indices of different ports.
The impedance matrix writes as $Z_{jj'} = -i \omega \sqrt{\frac{\mu}{\epsilon}} \sum_{em-modes} \frac{\omega_n \hat{c}_{nj'} \hat{c}_{nj}}{(\omega_n^2 - i\omega \gamma - \omega^2)}$

We next relate elements of the impedance matrix to the radiation resistance of the ports.
We first introduce the mode density.
We use $N(\omega)$ as the number of cavity modes with $\omega < \omega_n$.
Because we work with cavities with a high mode density, we have $\sum_{em-modes} \rightarrow \int d\omega_n N'(\omega_n), \quad N'(\omega_n) = dN(\omega)/d\omega |_{\omega = \omega_n}$.
The real part of the impedance $Re(Z_{jj'}) = -i \omega \sqrt{\frac{\mu}{\epsilon}} \int d\omega_n N' \frac{\omega_n \hat{c}_{nj'} \hat{c}_{nj}}{(\omega_n^2 - i\omega \gamma - \omega^2)} \approx  \omega \sqrt{\frac{\mu}{\epsilon}} \frac{\pi}{2} N' \hat{c}_{nj'} \hat{c}_{nj}$.


We redefine the coupling coefficient as $\sqrt{\frac{\mu}{\epsilon}} \hat{c}_{nj'} \hat{c}_{nj} = \left[ R_{rad}(\omega_n) / \left( \frac{\pi}{2} \omega_n N' \right) \right] c_{nj}^2$ where $c_{nj}$ is a zero mean, unit variance Gaussian random variable.
The quantity $R_{rad,j}$ is the radiation resistance of port $j$.
Thus we have defined the mode-port coupling coefficient as $c_{nj} = \left[ \sqrt{\frac{\epsilon}{\mu}} R_{rad, j}(\omega_n) / \left( \frac{\pi}{2} \omega_n N' \right) \right]^{1/2} \hat{c}_{nj}$.
We also have the system impedance matrix as
\begin{equation} \label{eq:f3}
    Z_{jj'} = -i \omega \sum_{em-modes} \frac{2(R_{rad,j}R_{rad,j'})^{1/2}}{\pi N'} \frac{c_{nj'} c_{nj}}{(\omega_n^2 - i\omega \gamma - \omega^2)}.
\end{equation}
Note that Eq. \ref{eq:f3} has the same form as the frequency domain single cavity RCM \cite{Gradoni2014}.

The final TD-RCM mode equations are
\begin{widetext}
\begin{equation} \label{eq:all4}
\begin{aligned}
    & \frac{\partial^2}{\partial t^2} U_n + \gamma_n \frac{\partial}{\partial t} U_n + \omega_n^2 U_n = - \left( \frac{\mu}{\epsilon} \right)^{1/4} \omega_n \sum_{ports} \left[ \frac{2R_{rad, j}(\omega_n)}{\pi \omega_n N'} \right]^{1/2} c_{nj} \frac{\partial}{\partial t} I_j\\
    & \gamma_n = \omega_n / Q_n \\
    & C \frac{\partial}{\partial t} V_{jc} = - I_j \\
\end{aligned} \quad.
\end{equation}
\end{widetext}
Here $N'(\omega_n)= 1/\Delta \omega_n$ is the inverse of the mode spacing at $\omega_n$, and $V_{jc}$ describes the electrostatic contribution.
The quantity $C$ is the port radiation reactance at lower frequencies as in the main text.
The TD-RCM port voltage equation is
\begin{equation} \label{eq:all5}
    V_j = - \left( \frac{\epsilon}{\mu} \right)^{1/4} \sum_{em-modes} \left[ \frac{2R_{rad, j}(\omega_n)}{\pi \omega_n N'} \right]^{1/2} c_{jn} U_n - V_{jc}.
\end{equation}
With Eqs. \ref{eq:all4} and \ref{eq:all5}, we have the complete equations of motion of TD-RCM.
The coefficient $K_{nj} = - \left( \frac{\mu}{\epsilon} \right )^{\frac{1}{4}} \left[ \frac{2R_{rad,j}(\omega_n) \, \Delta\omega_n}{\pi\omega_n } \right]^{\frac{1}{2}}$ is the combined coefficient on the RHS of the mode equation Eq. \ref{eq:all4}, and $K'_{jn}$ is the corresponding coefficient in Eq. \ref{eq:all5}.

\subsection{C. TD-RCM Treatment of Linear and Non-Linear Loads - Spectral Response}

\begin{figure*}
\centering
\includegraphics[width=1\textwidth]{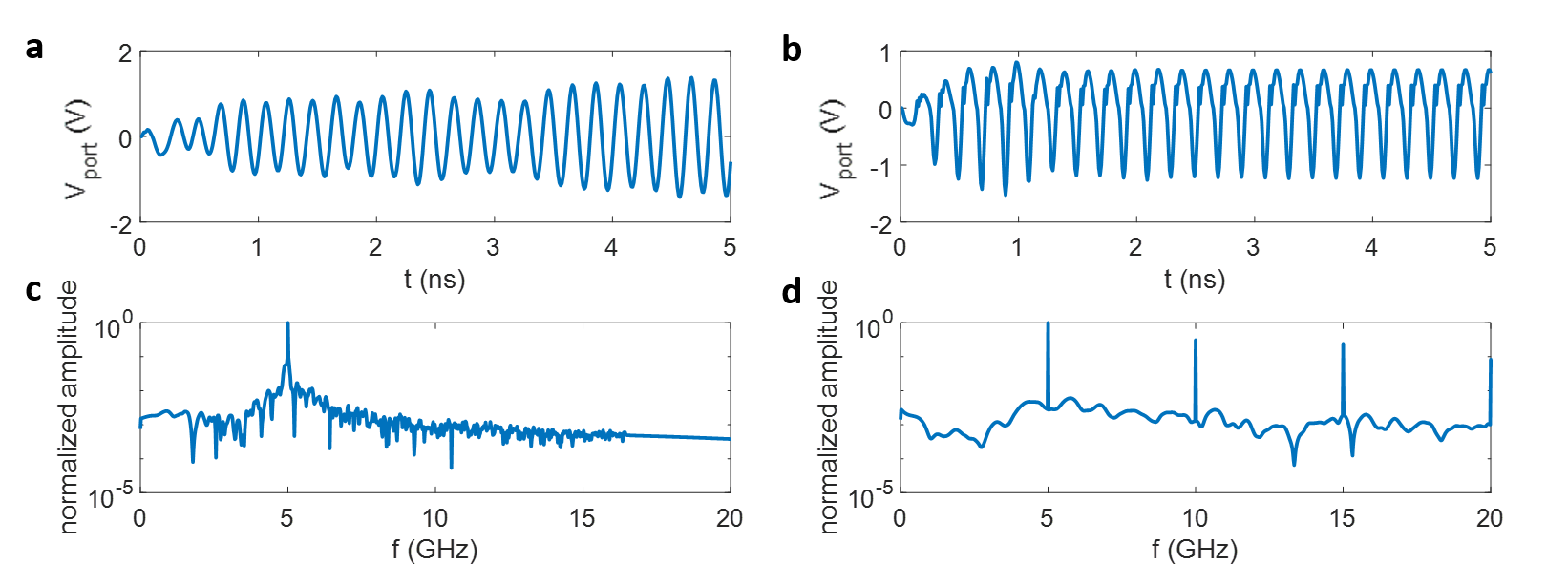}
\caption{\label{fig:nlin} 
The TD-RCM simulated RX port signals. (a) and (b) show the time-domain RX port signal $V_{port}$ for a 50 $\Omega$ linear load and a diode load, respectively. The Fourier transform of the time-domain waveforms in (a) and (b) are shown in (c) and (d).
}
\end{figure*}

To expand the applicability of the TD-RCM model, we have presented the treatment of nonlinear port loads in Section III.B.
Here we test the nonlinear capability of TD-RCM with a simple single-tone stimulation of a cavity with a nonlinear output port.
In this numerical exercise, we inject a single frequency 5 GHz sine-wave for 50 ns through a linear TX port into a model cavity with $M=2$ ports, and study the induced voltage signal at a second RX port in the cavity.
The loss parameter of the system is set at $\alpha = 1$, and $N=1000$ system modes are used.
The mean-mode-spacing is set as $\Delta k^2 = 4\pi/A = 109.3 m^{-2}$ where $A = 0.115 m^2$ is the two-dimensional billiard area.
For convenience, the frequency-independent radiation resistance of both ports is set to $(18 + 50 i)\Omega$ (measured at 5 GHz).
Because we are only testing the port load responses, we have fixed the system spectrum ($\omega_n$ in Eq. \ref{eq:modeV}) and the port-mode coupling coefficients  ($c_{jn}$ in Eqs. \ref{eq:modeV} and \ref{eq:portV}).
For the linear system test, the load connected at the RX port is a 50 $\Omega$ linear impedance (Fig. \ref{fig:portload} (a)).
For the nonlinear system test, the RX port is a diode-loaded port, as shown in Fig. \ref{fig:portload} (b).
The following parameters are used for the diode-loaded port ($R = 50\Omega, C = 1.5e^{-20}F, I_0 = 1e^{-7}A$, and $T = 300K$).
The amplitude of the CW input sine-wave is 5V.
The computed time-domain linear and nonlinear RX port results are shown in Figs. \ref{fig:nlin} (a) and (b), respectively.
One can already see hints of nonlinearity from Fig. \ref{fig:nlin} (b) by eye.
We next show the spectrum of the linear and nonlinear time-domain waveforms in Figs. \ref{fig:nlin} (c) and (d), respectively.
Here, the linear-port spectrum is dominated by one peak at the fundamental frequency (5 GHz), while the spectrum of the nonlinear-port shows a clear peak at the second and several higher harmonics.
Thus the TD-RCM is capable of simulating nonlinear elements and emulating proper nonlinear behaviors.
Note that the speed of TD-RCM computation will slow down when nonlinear elements are included.

\end{spacing}

\end{document}